\PassOptionsToPackage{pdfpagelabels=false}{hyperref}

\documentclass[preprint,authoryear,12pt]{elsarticle}  

\usepackage{newtxtext,newtxmath}

\usepackage[T1]{fontenc}
\usepackage{newtxtext,newtxmath}
\usepackage{ae,aecompl}
\usepackage{placeins}
\usepackage{graphicx}	
\usepackage{amsmath}	
\usepackage{gensymb}
\usepackage{natbib} 
\usepackage{subcaption}
\usepackage{longtable}
\PassOptionsToPackage{hyphens}{url}
\usepackage{hyperref}
\usepackage{multicol,multirow,makecell}
\newcommand{\vthirteen}{Video 1}
\newcommand{\vfourteen}{Video 2}
\newcommand{\vfifteen}{Video 3}
\newcommand{\vnineteen}{Video 4}
\newcommand{\vtwenty}{Video 5}
\newcommand{\vtwentyone}{Video 6}
\newcommand{\vtwentytwo}{Video 7}
\newcommand{\vtwentyfour}{Video 8}
\newcommand{\vtwentyfive}{Video 9}
\newcommand{\vsixteen}{Video 10}
\newcommand{\vseventeen}{Video 11}
\newcommand{\veightteen}{Video 12}

\title{Ricochets on Asteroids II: 
Sensitivity of laboratory experiments of low velocity grazing impacts on substrate grain size}

\author[a1]{Esteban Wright\corref{cor1}} 
\ead{ewrig15@ur.rochester.edu}
\author[a1]{Alice C. Quillen}
\ead{alice.quillen@rochester.edu}
\author[a2]{Paul S\'anchez}
\ead{diego.sanchez-lana@colorado.edu}
\author[a4,a5]{Stephen R. Schwartz}
\ead{srs51@email.arizona.edu}
\author[a1,a5]{Miki Nakajima}
\ead{mnakajima@rochester.edu}
\author[a3]{Hesam Askari}
\ead{askari@rochester.edu}
\author[a3]{Peter Miklavcic}
\ead{pmiklavc@ur.rochester.edu}

\address[a1]{Department of Physics and Astronomy, University of Rochester, Rochester, NY 14627, USA}
\address[a2]{Colorado Center for Astrodynamics Research, The University of Colorado Boulder, UCB 431, Boulder, CO 80309-0431, United States}
\address[a3]{Department of Mechanical Engineering, University of Rochester, Rochester, NY 14627, USA}
\address[a4]{Lunar and Planetary Lab, University of Arizona, Tucson, AZ, USA}
\address[a5]{Laboratoire Lagrange, Universit\'e C\^ote d'Azur, Observatoire de la C\^ote d'Azur, CNRS, C.S. 34229, 06304 Nice Cedex 4, France}

\cortext[cor1]{Corresponding author}

\begin{document}

\begin{abstract}
We compare low velocity impacts that ricochet with the same impact velocity and impact angle into granular media with similar bulk density, porosity, and friction coefficient but different mean grain size. The ratio of projectile diameter to mean grain length ranges from 4 in our coarsest medium to 50 in our finest sand.
Using high speed video and fluorescent markers, we measure the ratio of pre- to post-impact horizontal and vertical velocity components, which we refer to as coefficients of restitution, and the angle of deflection caused by the impact in the horizontal plane.  
Coefficients of restitution are sensitive to mean grain size with the ratio associated with the horizontal velocity component about twice as large for our coarsest gravel as that for our finest sand. 
This implies that coefficients for hydro-static-like, drag-like and lift-like forces, used in empirical force laws, are sensitive to mean grain size.  The coefficient that is most strongly sensitive to grain size is the lift coefficient which decreases by a factor of 3 between our coarsest and finest media.
The deflection angles are largest in the coarser media
and their size approximately depends on grain size to the 3/2 power.  
This scaling is matched with a model where momentum transfer takes place via collisions with individual grains. 
The dependence of impact mechanics on substrate size distribution should be considered in future models for populations of objects that impact  granular asteroid surfaces. 

\end{abstract}


\maketitle

\section{Introduction}

Laboratory study of impacts of marbles into sand showed that low velocity (a few m/s) projectiles can ricochet off granular materials, such as sand, at grazing angles (from the surface) up to $40\degree$
\citep{Wright_2020b}.
These laboratory experiments were done at 1\ g, however these and normal impact experiments can be scaled to low surface gravity conditions on astronomical bodies using dimensionless parameters (e.g., \citealt{goldman08,murdoch17}).
This suggestion builds upon a related body of work that has developed scaling relations for impact craters \citep{holsapple93}.

A dimensionless parameter used to predict the outcome (whether it would roll out, or ricochet out of its impact crater) of a low velocity grazing incidence projectile on the surface of a rubble asteroid is the Froude number or acceleration parameter
\begin{equation}
Fr  \equiv \frac{v_{impact}}{\sqrt{g R_p}},
\label{eqn:Fr}
\end{equation}
\citep{Wright_2020b}
where $v_{impact}$ is the projectile velocity at impact, $R_p$ is the projectile radius and $g$ is the gravitational acceleration. The Froude number is inversly related to the dimensionless $\pi_2$ parameter used by \citet{holsapple93}, which is the ratio of the lithostatic pressure to the impact pressure of the projectile.
Crater scaling laws
\citep{holsapple93,ambroso05} and empirical force models for normal impacts into granular media \citep{tsimring05,katsuragi07,goldman08,katsuragi13}
also depend upon the ratio of projectile to substrate bulk density ${\rho_p}/{\rho_s}$.   
Laboratory studies have carried out normal impact experiments into a variety of granular media, however, differences in the force laws or penetration depth are usually attributed to variations in velocity, gravitational acceleration and projectile and substrate densities, rather than to the substrate grain properties such as grain size or shape or friction coefficients (e.g., \citealt{ambroso05,goldman08,katsuragi13}),
with an exception being \cite{ballouz21}. 

Impact experiments have previously shown sensitivity to the grain characteristics. For example, acceleration fluctuations during projectile penetration of normal impacts were interpreted as due to creation and annihilation of elements in the force chain network that are sensitive to size and shape of the grains \citep{goldman08,kondic12,bester19}.  The large contrast in bulk density of the granular media used in normal impact experiments (e.g., glass or bronze spheres, aluminum shot, or millet for \citealt{goldman08}; glass beads, rice, beach sand and sugar for  \citealt{katsuragi13}) made it difficult to differentiate between the role of substrate density and the role of other substrate properties such as friction coefficient, particle size or shape.  In most settings, differences in quantities such as penetration depth were attributed to substrate density \citep{ambroso05,goldman08,katsuragi13}.

Recent missions have shown that rubble pile asteroids' surfaces are geomorphologically diverse and heterogeneous \citep{veverka01,  miyamoto07,Lauretta_2019,DellaGiustina19,Michikami_2019}.  
Asteroids surfaces can be characterized by the size distribution of boulders, pebbles and particles seen in images.  Comparative imaging studies characterize the boulder size distribution with a power law index $\alpha$ with $\alpha= -2.65$ on (162173) Ryugu,
-3.05 on NEA (25143) Itokawa and 
-3.25 on (433) Eros \citep{Michikami2021}.
The size frequency distribution of boulders on (101955) Bennu is $-2.9 \pm 0.3$ \citep{DellaGiustina19}.

The number density of boulders (the number of boulders per square km) differs between these bodies.  
The number density of 30 m long boulders on Itokawa is about an order of magnitude larger than that on Eros and that on Ryugu and Bennu is a few times larger than that on Itokawa (see Figure 14 by \citealt{Michikami2021} and Figure 3 by \citealt{DellaGiustina19}).

Size frequency boulder distributions on asteroid surfaces also depend on location on the surface.
On Ryugu \citep{Michikami_2019} and on Bennu \citep{DellaGiustina19} there is a paucity of large boulders in the equatorial region. 
Itokawa displays smooth regions, comprised of cm sized grains, and rough regions, comprised of boulders with sizes of 10s of meters \citep{miyamoto07}. Eros notably has remarkably flat regions known as ponded deposits which contain fine particles less than $\sim 30$ cm in size \citep{cheng02}.

Re-accumulation after a large collision \citep{michel13,walsh19}, mass-shedding \citep{hirabayashi15,Scheeres_2015}, ejecta from impacts \citep{Wada_2006,wright19}, and particle ejection events \citep{Bottke_2020,Chesley_2020}, can give a population of low velocity particles that can return to impact the asteroid surface at low velocity.  These particles can bounce or ricochet off the surface \citep{Chesley_2020}.
The heterogeneity of rubble found on astronomical objects and differences in size distributions on different regions of astronomical bodies motivates understanding how impacts are sensitive to the properties of the granular substrate.
The sensitivity of low velocity projectile behavior to the size of particles in the granular substrate is the focus of this paper.

In this paper we carry out laboratory experiments of a spherical projectile, a marble, launched at low velocity 
and at grazing impact angle into different granular media.  We compare impacts with the same projectile, impact velocity and angle. 
We restrict our substrates to similar density rocky materials and we use a glass projectile so as to approximately match the projectile and grain material densities. The granular media differ primarily in the mean size of the particles and our goal is to probe 
the sensitivity of the impact mechanics to mean grain size.
Nomenclature used throughout this paper is given in Table \ref{tab:nomen}.

\begin{table}
{
\caption{Nomenclature}
\label{tab:nomen}
\begin{tabular}{ll}
\hline
Gravitational acceleration & $g$ \\
Projectile mass & $M_p$  \\
Projectile radius & $R_p$ \\
Projectile density & $\rho_p$ \\
Grain mass & $m_g$ \\
Granular substrate mean density & $\rho_s$ \\
Grain semi-major axis & $a_g$ \\
Grain semi-middle axis   & $b_g$ \\
Grain semi-minor axis & $c_g$ \\
Projectile-grain size ratio & $\pi_{grain}= R_p / \bar a_g$ \\
Projectile velocity vector & ${\bf v}$ \\
Projectile velocity at impact & $v_{impact}$ \\
Impact angle & $\theta_{impact}$ \\
Projectile angular acceleration &  $\dot{\omega}$ \\
Deflection angle in xy-plane & $\alpha$ \\
Standard deviation of deflection angle & $\sigma_\alpha$ \\
Porosity (substrate) & $\phi$ \\
Coefficient of static friction  & $\mu_s$ \\
Angle of repose & $\theta_r$ \\
Coefficients of Restitution & $e_x, e_z$ \\
Standard deviation of restitution coefficients & $\sigma_{e_x}, \sigma_{e_z}$ \\
Effective friction coefficient & $\mu_{eff}$ \\
Standard deviation of $\mu_{eff}$ & $\sigma_{\mu_{eff}}$ \\
Froude number  &  Fr $= \frac{v_{impact}}{\sqrt{R_p  g}}$  \\
Lift coefficient & $C_L$ \\
Drag (horizontal) coefficient & $C_{Dx}$ \\
Time to exit impact region & $t_e$ \\
\hline
\end{tabular}\\

For a normal impact $\theta_{impact} = 90\degree$.
The vertical $z$ coordinate is positive upward.
The horizontal $x$ coordinate is positive in the initial direction of projectile motion.}
\end{table}

\section{Laboratory experiments of oblique impacts into granular substrates of different mean grain size}
\label{sec:exp}

We carry out low velocity grazing impact experiments similar to those described by \cite{Wright_2020b}, except we use four granular substrates with different mean grain sizes.  We use sieves to restrict the grain size range in each substrate.   Photo-scans are then used to characterize the size and shape distribution in each substrate. Two sets of experiments are presented: one set for the projectile dynamics pre and post ricochet, and the second set for the deflection of the projectile away from the impact point.

\begin{figure}
\centering
\begin{subfigure}{\linewidth}
   \centering
   \includegraphics[width=0.6\linewidth,scale=1,trim = {0 0 0 0}, clip]{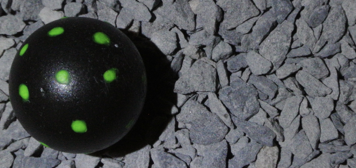}
   \caption{Coarse gravel $\pi_{grain} \approx 4$}
   \label{fig:grav_coarse_pi} 
\end{subfigure}
\begin{subfigure}{\linewidth}
   \centering
   \includegraphics[width=0.6\linewidth,scale=1,trim = {0 0 0 0}, clip]{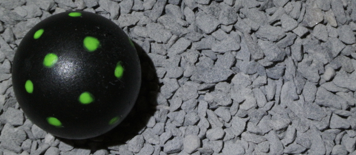}
   \caption{Fine gravel $\pi_{grain} \approx 5$}
   \label{fig:grav_fine_pi}
\end{subfigure}
\begin{subfigure}{\linewidth}
   \centering
   \includegraphics[width=0.6\linewidth,scale=1,trim = {0 0 0 0}, clip]{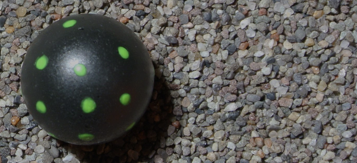}
   \caption{Dark sand $\pi_{grain} \approx 17$}
   \label{fig:sand_dark_pi}
\end{subfigure}
\begin{subfigure}{\linewidth}
   \centering
   \includegraphics[width=0.6\linewidth,scale=1,trim = {0 0 0 0}, clip]{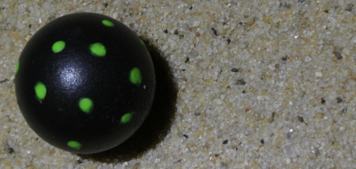}
   \caption{Light sand $\pi_{grain} \approx 51$}
   \label{fig:sand_light_p}
\end{subfigure}
\caption{Images of the four different granular substrates and the marble projectile for comparison with the grain size. The marble has a diameter of 16.15 mm. The properties of the granular substrate, including the mean semi-major axis of the grains $\bar a_{g}$ used to calculate $\pi_{grain}$, are listed in Table \ref{tab:granular}
}
\label{fig:grain_img}
\end{figure}

\begin{table*}
{
\caption{Granular Properties}
\label{tab:granular}
\resizebox{\textwidth}{!}{
\begin{tabular}{lllllllll}
\hline
Material & \makecell{Mean semi-\\major axis} & \makecell{Mean semi-\\middle axis} &  \makecell{Marble-grain \\ length ratio} & \makecell{Grain \\ Density} & \makecell{Substrate (bulk) \\ Density} & Porosity & \makecell{Angle of \\ Repose} & \makecell{Static friction \\ coefficient} \\
 & ${\bar a}_g$ (mm) & ${\bar b}_g$ (mm) & $\pi_{\rm grain} = R_p/ {\bar a}_g$ & $\rho_g$ (g/cm$^3$) & $\rho_s$ (g/cm$^3$) & $\phi$ & $\theta_r$ & $\mu_s$ \\
\hline
Gravel, coarse & 2.18 & 1.41 & 3.7 & 2.61 & 1.33 & 0.491 & 40\degree & 0.84\\
Gravel, fine & 1.61 & 0.95 & 5.0 & 2.58 & 1.31 & 0.492 & 39\degree & 0.81\\
Sand, dark & 0.49 & 0.36 & 16.5 & 2.50 & 1.43 & 0.428 & 36\degree & 0.73\\
Sand, light & 0.16 & 0.12 & 50.5 & 2.40 & 1.52 & 0.366 & 40\degree & 0.84\\
\hline
\hline
\end{tabular}
}
Notes: The coefficient of static friction for the granular material is computed from its angle of repose $\mu_s = {\rm tan}(\theta_r)$.
The values for the mean semi-major and semi-middle axes were computed using the size distributions shown in Figure \ref{fig:size_dist}.} 
\end{table*}

\subsection{Grain size and shape distributions}
\label{sec:size_dist}

To mimic an asteroid environment,  our granular media were chosen to have rocky materials similar to that of our projectile (glass). 
We use a gravel comprised of gray shale and two different sands as substrates.  
The gravel used is available for road and driveway base layers in upstate New York.
We sieved the gravel into two separate size ranges.  For the coarser gravel, grains had a size range of 2-3 mm. The finer gravel had grains with a size range of 1-2 mm.
We use two different sands: black beach sand, and light playground sand. The black beach sand was obtained from the Lake Ontario shore and it is coarser than the playground sand. 
The black sand was sieved to a size range of 0.5-1 mm.
The light sand is our smallest grain size substrate with a grain size of >0.5 mm. 

We describe the shape of a grain with a tri-axial ellipsoid with semi-axes $a_g$, $b_g$ and $c_g$. Here $a_g\ge b_g \ge c_g$ so $a_g$ is the long axis. Our sieves have square mesh holes.
A sieve with 1 mm wide holes would allow particles with $2b_g < 1$ mm and $2c_g < 1$ mm to pass through it.  As long as these conditions on the semi-middle and semi-minor axis are obeyed, this same sieve could let long particles with $2a_g >1$ mm pass through it.

Photographs of the four granular media are shown in Figure \ref{fig:grain_img} along with the marble projectile for scale.
To characterize the size scale of our projectile and substrate we define a dimensionless parameter
\begin{equation}
    \pi_{grain} \equiv \frac{R_p}{{\bar a}_g}
\end{equation} 
where $R_p$ is the projectile radius and ${\bar a}_g$ is the mean semi-major axis length. 
The values for the mean semi-major axis for each substrate and its associated $\pi_{grain}$ value 
are listed in Table \ref{tab:granular}.

\begin{figure}
\centering
\includegraphics[width=0.5\columnwidth]{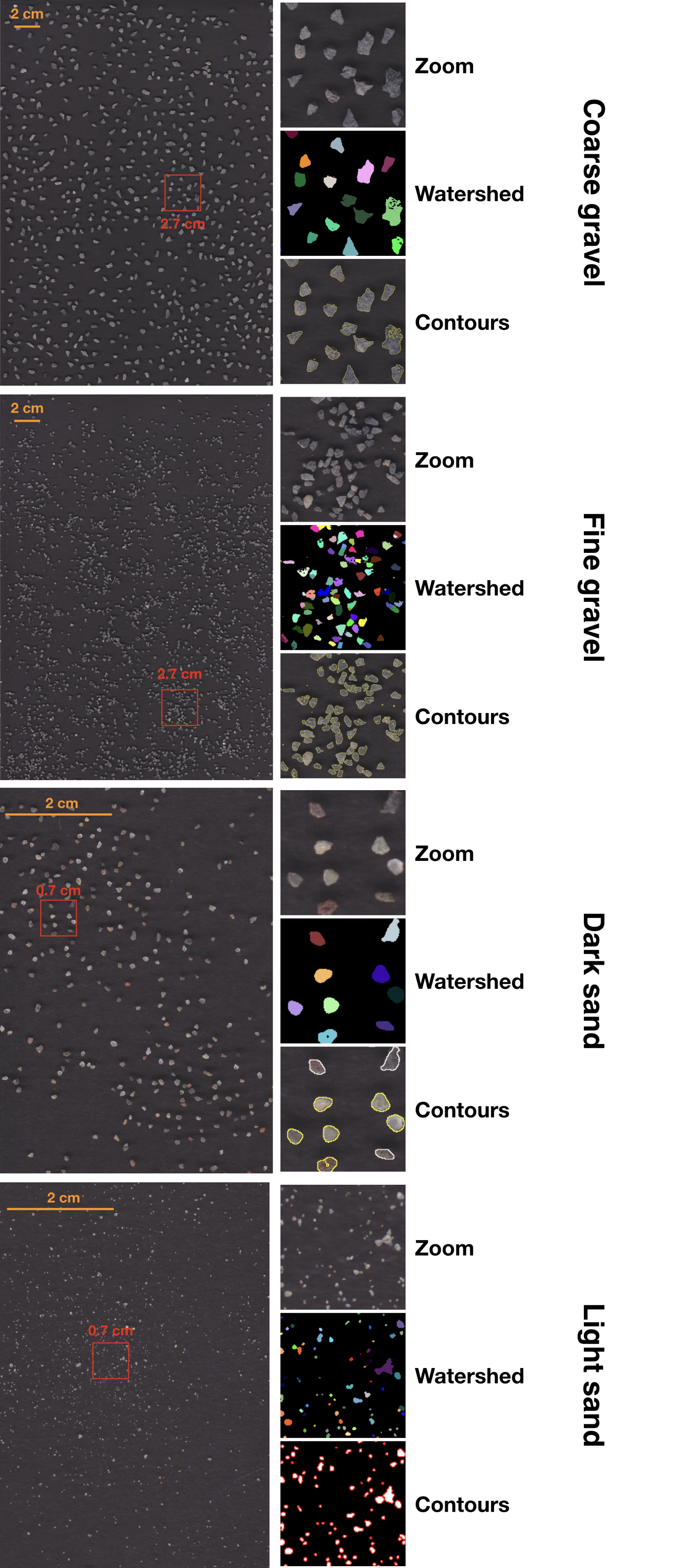}
\caption{On the left are the photo scanned images of grains from each substrate.
From top to bottom we show the coarse gravel, the fine gravel, the dark sand, and the finest light sand.
Each image was scanned with a resolution of 600 dot per inch.
In each image we zoom into regions enclosed by the red square.
The coarse and fine gravels have a red square which
measures about 2.7 cm in width.
For the light and dark sand the red square is about 0.7 cm.
The squares are different physical lengths since the images for the dark and light sand have been enlarged to show greater detail of the individual grains.
To the right and stacked vertically are three images of the zoomed region.
The top image is a zoomed image of this portion from the original.
The middle images show the resulting colored blobs from the watershed technique. 
The bottom images show the contours found from the watershedding algorithm.
Measurements of the contour's semi-major and semi-middle axes were used to generate the size distributions in Figure \ref{fig:size_dist}
and axis ratio distributions in Figure \ref{fig:axis_ratio_dist}.
The number of contours found in each full image for the coarse gravel was 725, 3511 for the fine gravel, 
285 for the dark sand, and 825 for the light sand.
\label{fig:watershed}
}
\end{figure}

To measure the grain size distribution in each of our four substrates we take multiple images of grains 
using a Canon CanoScan LiDE 100 photo scanner with a black background for contrast.
The images were scanned with a resolution of 600 dots per inch (dpi).
The photo scanner provides uniform lighting and minimizes shadows cast from the grains.   
To measure grain sizes and shapes,  we use the open source image analysis software Fiji (ImageJ) \citep{fiji,imagej} and the Interactive Watershed plugin written and maintained by Benoit Lombardot \citep{vincent91,najman96,lotufo2000}.
Watershed is an image analysis technique which finds the local minimum intensities in a gray-scale image, treating these minimums as a source and filling the local 'basin' until it contacts a basin from another source forming the contour. This technique gives contours for each grain, and is effective at identifying individual grains even when two are in contact.
Example of the photo-scanned images, the watershed mask, and the shape contours for the individual grains are shown in Figure \ref{fig:watershed}.
The apparent axis lengths of each grain are measured from their respective contour.

\begin{figure*}
\centering
\begin{subfigure}{0.475\linewidth}
   \centering
   \includegraphics[width=\linewidth,trim = {0 0 0 0}, clip]{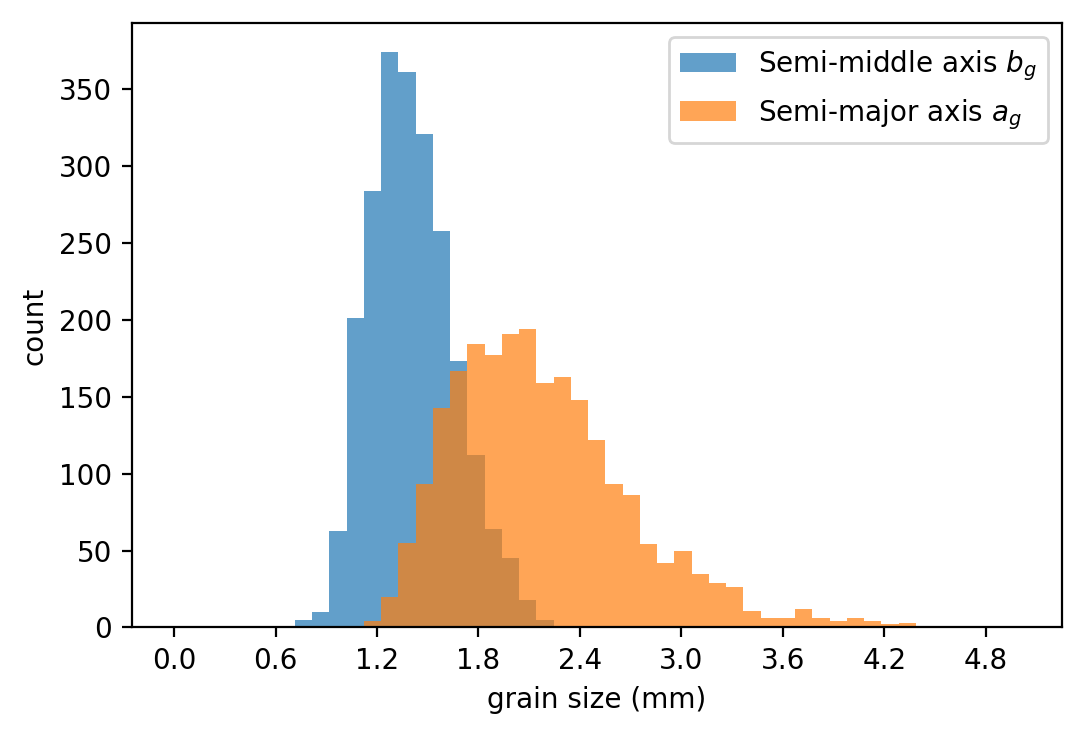}
   \caption{Coarse gravel}
   \label{fig:gravel_large_dist}
\end{subfigure}
\begin{subfigure}{0.475\linewidth}
   \centering
   \includegraphics[width=\linewidth,trim = {0 0 0 0}, clip]{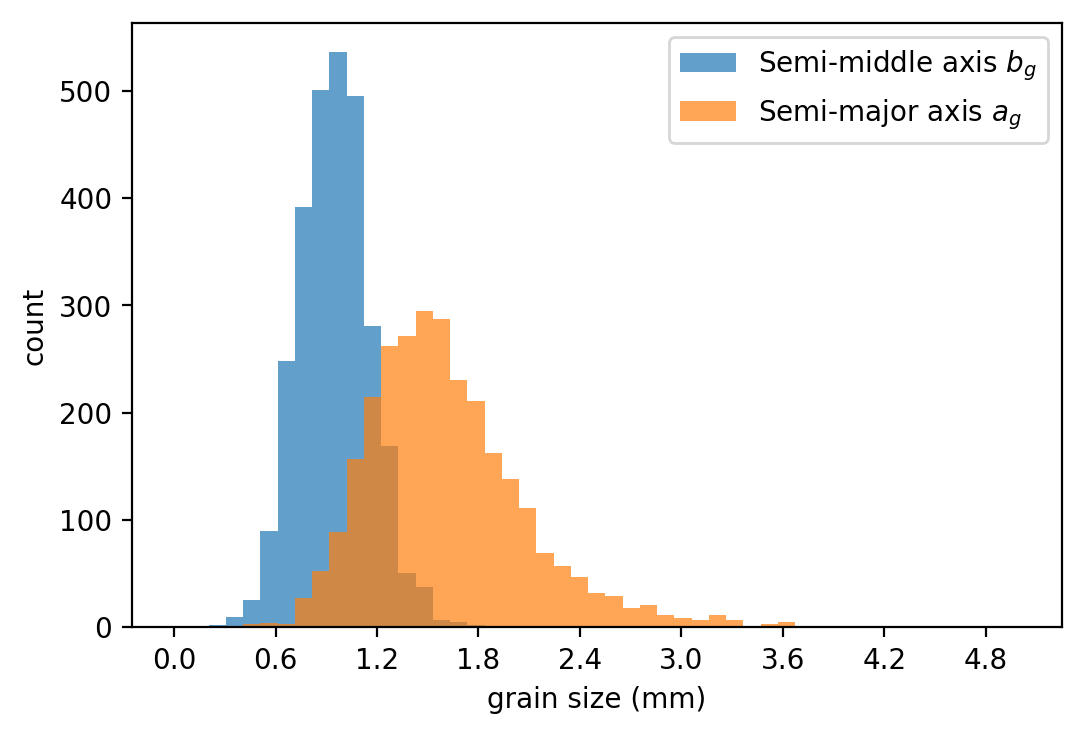}
   \caption{Fine gravel}
   \label{fig:gravel_mid_dist}
\end{subfigure}
\begin{subfigure}{0.475\linewidth}
   \centering
   \includegraphics[width=\linewidth,trim = {0 0 0 0}, clip]{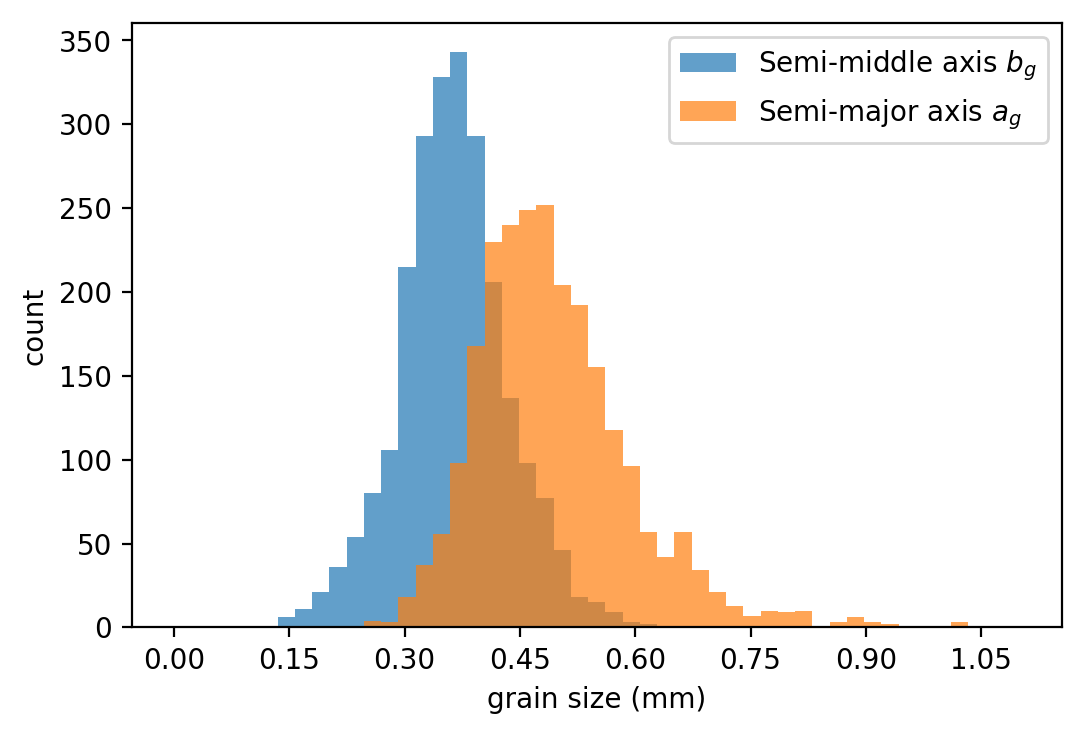}
   \caption{Dark sand}
   \label{fig:sand_dark_dist}
\end{subfigure}
\begin{subfigure}{0.475\linewidth}
   \centering
   \includegraphics[width=\linewidth,trim = {0 0 0 0}, clip]{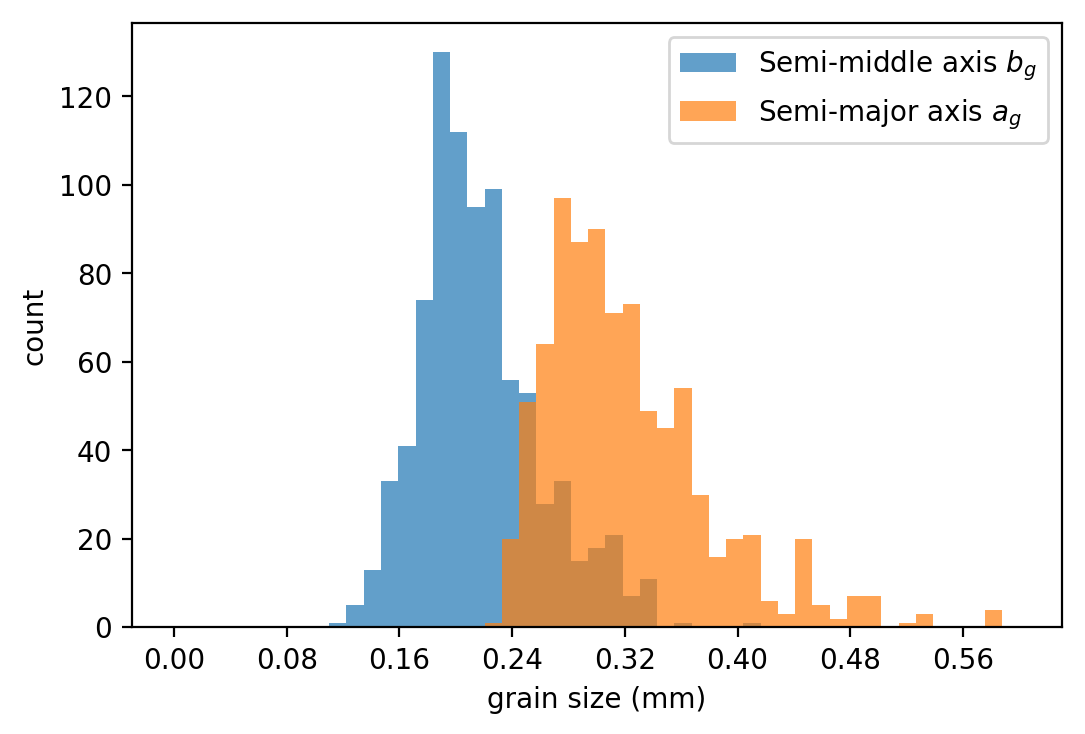}
   \caption{Light sand}
   \label{fig:sand_light_dist} 
\end{subfigure}
\caption{Distributions of the semi-major and semi-middle axes for all four grain sizes.
The distributions were measured from images of grains taken with a photo scanner and analyzed with Fiji (ImageJ) plugin \texttt{Interactive Watershed}, a watershed algorithm to find the boundaries of individual grains. The grains' semi-major and minor axes were then measured from the contours.
}
\label{fig:size_dist}
\end{figure*}

We assume that all grains lie flat on the scanner surface and so we can measure the semi-major and semi-middle axis lengths $a_g,b_g$ but we would not be able to measure the smallest axis length $c_g$ as this axis is oriented perpendicular to the scanner plane.  
We interpret the semi-major axis and semi-middle axes of the contour for a single grain to approximately be equal to the grain's semi-major $a_g$ and semi-middle axes $b_g$.  
The axis lengths are converted to physical units using a scale bar that we placed in each image.
To eliminate dust and noise in the black background, contours with area below 0.18 mm$^2$ (100 px) were discarded for the gravel and coarser sand. 
For the light sand, contours with area less than 0.04 mm$^2$ (20 px) were removed. 
The semi-major and semi-middle axis grain size distributions are shown in Figure \ref{fig:size_dist} for all four of our substrates.  

\begin{figure}
\centering
\includegraphics[width=\columnwidth,trim =  0 0 0 0, clip]{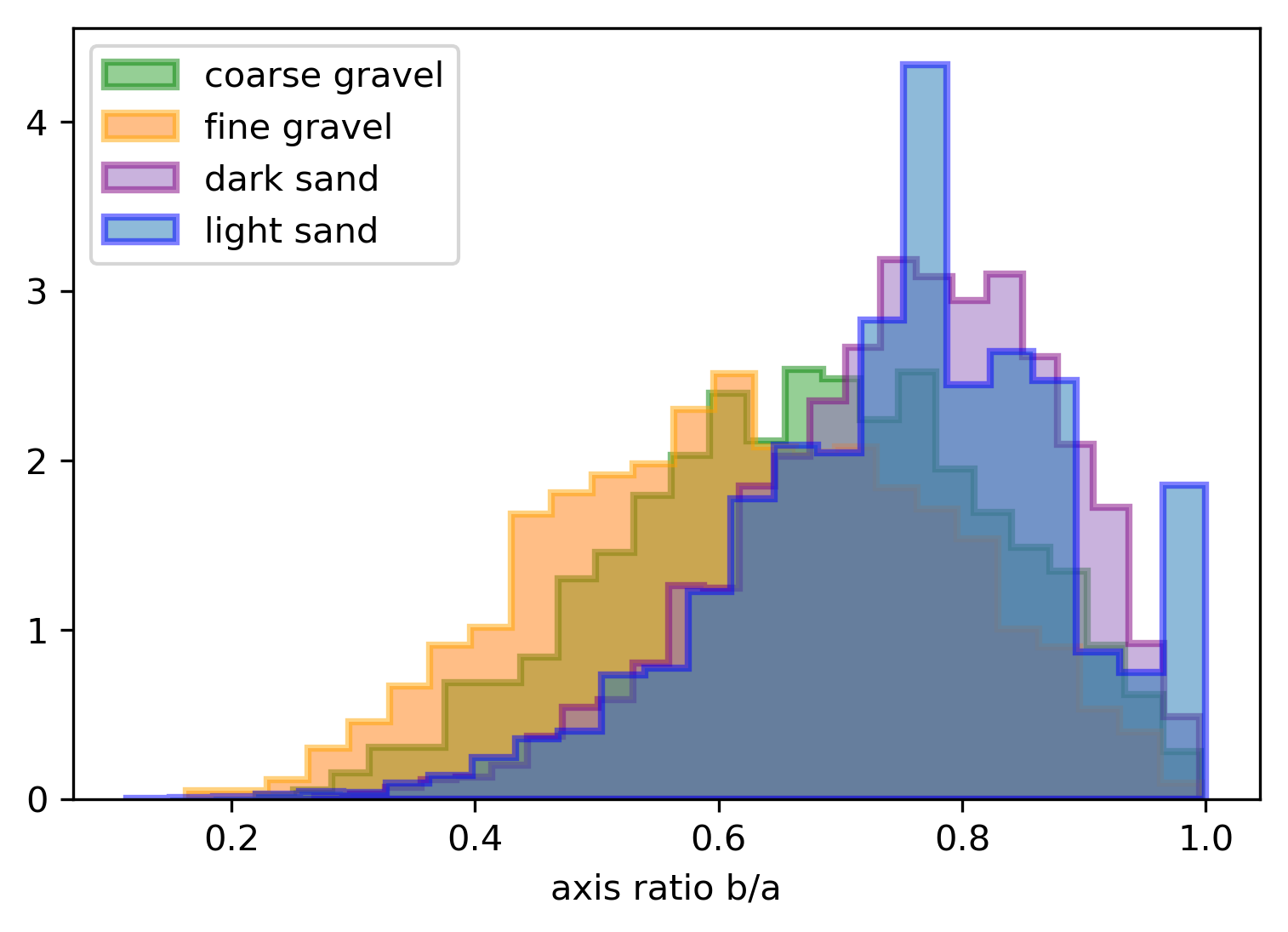}
\caption{Major-middle axis ratio $a_g/b_g$ distributions for the 4 granular media. The histograms are normalized such that they integrate to one. The gravel grains tend to be more elongated than the sand grains. 
\label{fig:axis_ratio_dist}
}
\end{figure}

Using the apparent semi-axes for each grain
we compute the axis ratio $a_g/b_g$ for each grain and plot the axis ratio or shape distributions for each substrate in Figure \ref{fig:axis_ratio_dist}.
This figure shows that the axis ratios are similar in the four media. 
The gravel particles tend to be more elongated than the sand particles. 
Shale is fissile and laminated, so when crushed, it tends to split into thin pieces.

\subsection{Properties of the granular media}
\label{sec:gran_prop}

We measured the bulk substrate density $\rho_s$ and grain density $\rho_g$ for each of our substrates. 
For bulk density $\rho_s$ we measured the mass of each substrate in a 1000 mL volume.
To estimate 
grain density $\rho_g$, we measured the volume of water needed to fill the air voids of the dry substrate in a 1000 mL volume.
Dividing the dry mass of the grains by the grain volume gives the grain density $\rho_g$.
We estimate the porosity of our substrate by taking the ratio of the bulk substrate and grain densities.
The values for substrate mean and granular densities and porosities are listed in Table \ref{tab:granular}.  
The porosity is highest in the coarse gravel and lowest in the light sand.  The grain densities in the four substrates are similar to each other and that of the projectile.  The coarse gravel has
the highest grain density which is 13\% higher than that of the glass marble projectile with a density of  $\rho_p = 2.53$ g cm$^{-3}$.

The angle of repose for each size was measured from the maximum exterior slope of a mound of material. 
We calculate a static friction coefficient $\mu_s$ of the granular media by taking the tangent of the angle of repose.
The substrate angles of repose and friction coefficients for our granular media are given in Table \ref{tab:granular}.
All of our granular media had similar angles of repose and associated static friction coefficients.

\begin{figure}
\includegraphics[width=3.5in]{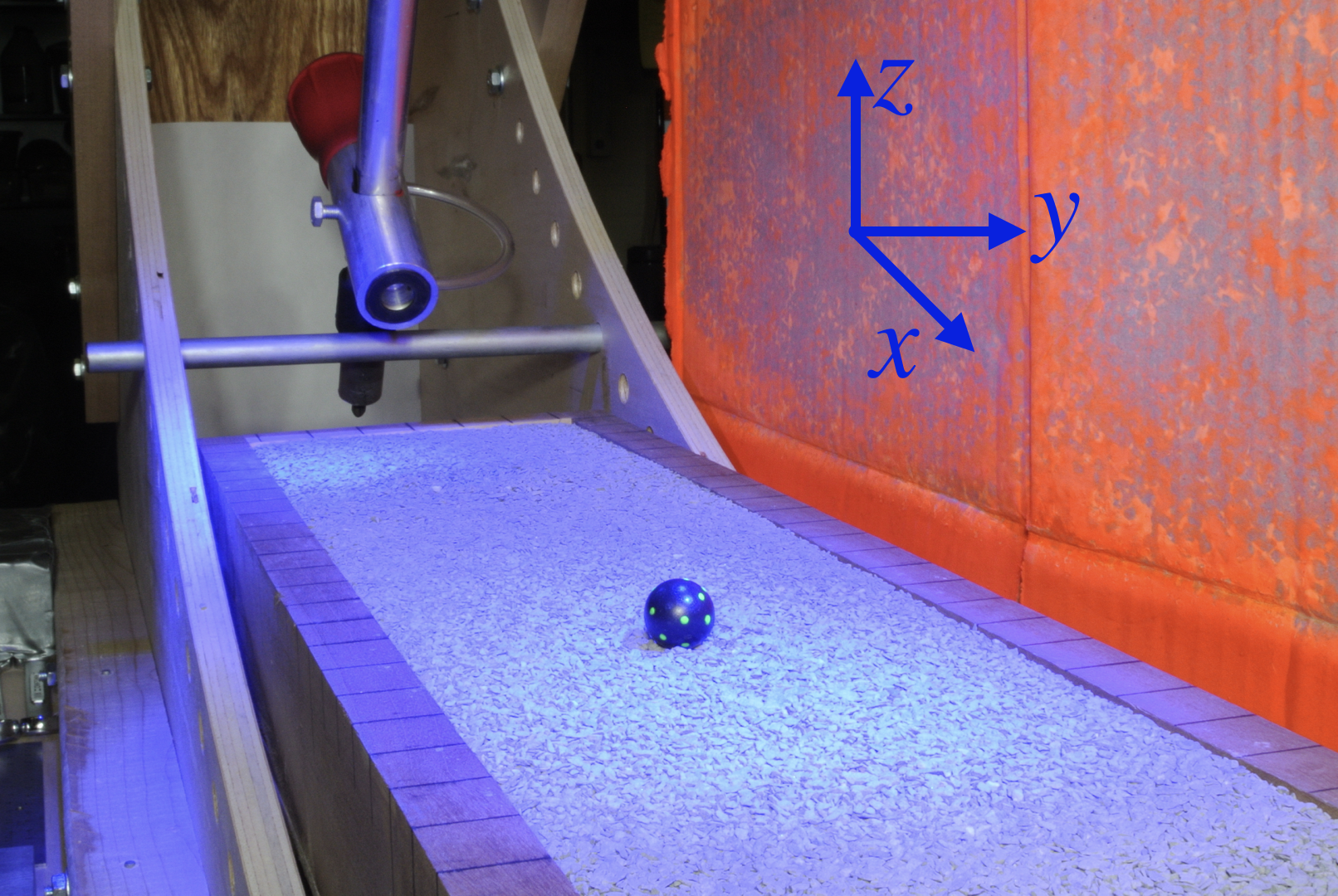}
\caption{A close up image of the pendulum arm and the marble projectile. The z-axis is vertical and the positive y-axis is towards the orange background.
The x-axis is in the direction of marble travel (i.e. along the length of the tray).
The marble is launched from the aluminum tube and is held in place by a small amount of suction from the red turkey baster. The marble is launched when the pendulum strikes the horizontal stop-bar that sets the projectile's impact angle.
The white dot reflection is used to track the trajectory of the marble and the green dots are used to track the angular velocity of the marble.
The black lines on the edges of the tray are spaced by 2 cm.
\label{fig:setup_close}}
\end{figure}

\begin{figure}
\includegraphics[width=3.5in]{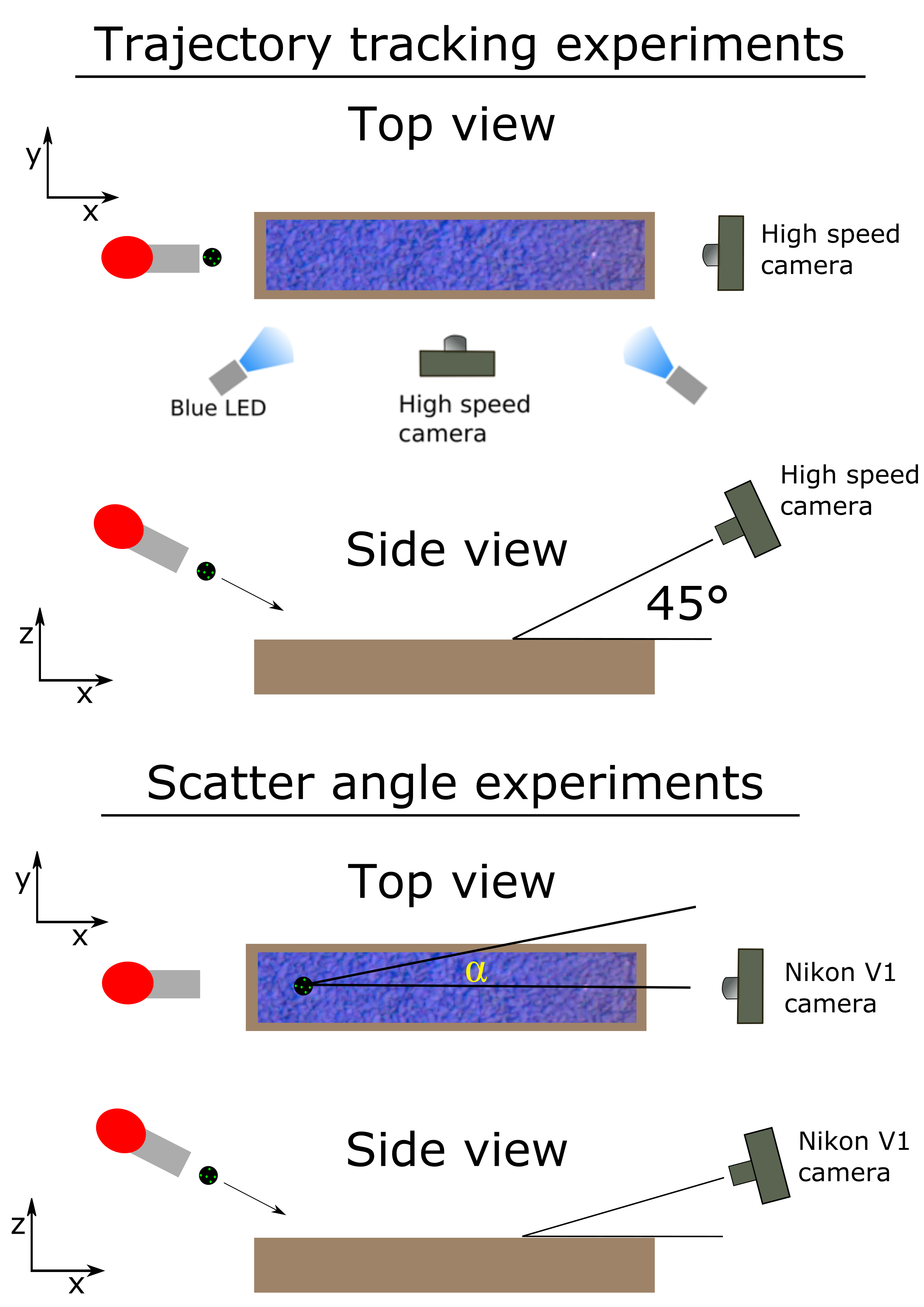}
\caption{Figure of the setup for oblique impact experiments. 
See Figure \ref{fig:setup_close} for the coordinate system.
In both images the projectile is launched from our pendulum arm (grey rectangle with red oval) that holds the projectile in place with a small amount of suction.
The top image is of the projectile tracking experiments used to measure the marble's trajectory, velocities, accelerations, and angular velocity (spin) in time. Two high speed cameras record the experiment perpendicular to and along the marble's travel direction (i.e. along the x-axis). 
The high speed camera looking along the length of the tray is elevated $45\degree$ above the horizontal surface.
Blue LEDs are used to illuminate fluorescent dots painted on the marble.
The bottom image is the experimental setup for measuring the deflection of the projectile from center. The deflection angle $\alpha$ is measured from the center of the tray to the marble's final position in its trajectory. One regular speed camera (Nikon V1) was directed along the length of the tray, slightly above the horizontal surface, and pointed down at an arbitrary angle.
\label{fig:setup_cartoon}}
\end{figure}

\subsection{Impact Experiments}

Our experiments use the same experimental apparatus as we described previously and used
to study low velocity ricochets \citep{Wright_2020b}.
A pendulum arm is used to launch a projectile into the substrate at a desired impact velocity and impact angle.
Figure \ref{fig:setup_close} shows the pendulum launcher in the background, the marble projectile in the foreground, and a coordinate system for the experiments in blue to the right. 

The projectile is a glass marble with a mass of 5.57 g and diameter of 16.15 mm.
The marble is launched from the pendulum head where it is held in place by a small amount of suction provided by the red turkey baster. 
The marble is launched when the pendulum strikes a horizontal stop-bar that sets the projectile's impact angle.
Marble properties are listed in Table \ref{tab:quantities}

All experiments discussed in this paper were launched with an impact angle of about 30\degree and impact velocity of $\sim$ 3.8 m/s.
The granular substrate was held in wooden trays with a depth of about 6 cm.
Three trays were used to hold substrate material. The tray lengths, widths and depths are listed in Table \ref{tab:quantities} along with the material characteristics of the particles
forming the substrate the tray held.

\begin{figure}
\centering
\begin{subfigure}[b]{0.95\linewidth}
   \centering
   \includegraphics[width=1\linewidth]{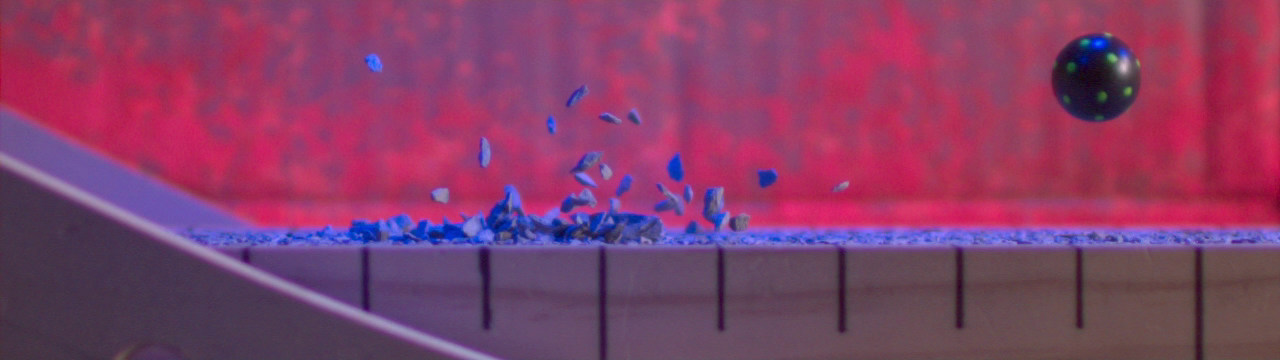}
   \caption{Coarse gravel, \vthirteen}
   \label{fig:impact_grav_large} 
\end{subfigure}
\begin{subfigure}[b]{0.95\linewidth}
   \centering
   \includegraphics[width=1\linewidth]{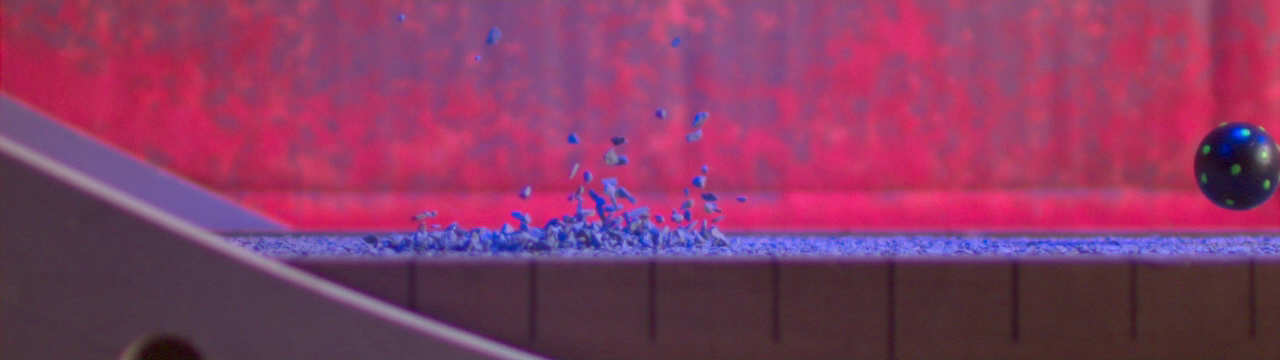}
   \caption{Fine gravel, \vnineteen}
   \label{fig:impact_grav_mid}
\end{subfigure}
\begin{subfigure}[b]{0.95\linewidth}
   \centering
   \includegraphics[width=1\linewidth]{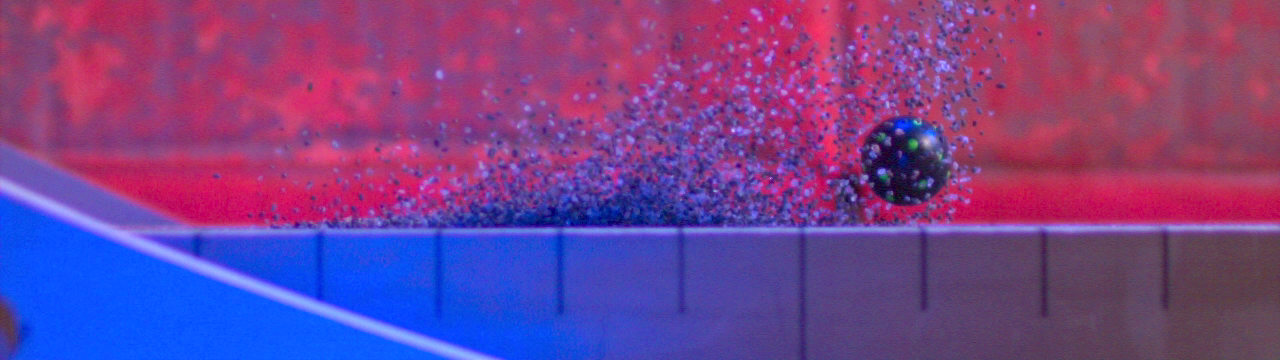}
   \caption{Dark sand, \vtwentytwo}
   \label{fig:impact_sand_dark}
\end{subfigure}
\begin{subfigure}[b]{0.95\linewidth}
   \centering
   \includegraphics[width=1\linewidth]{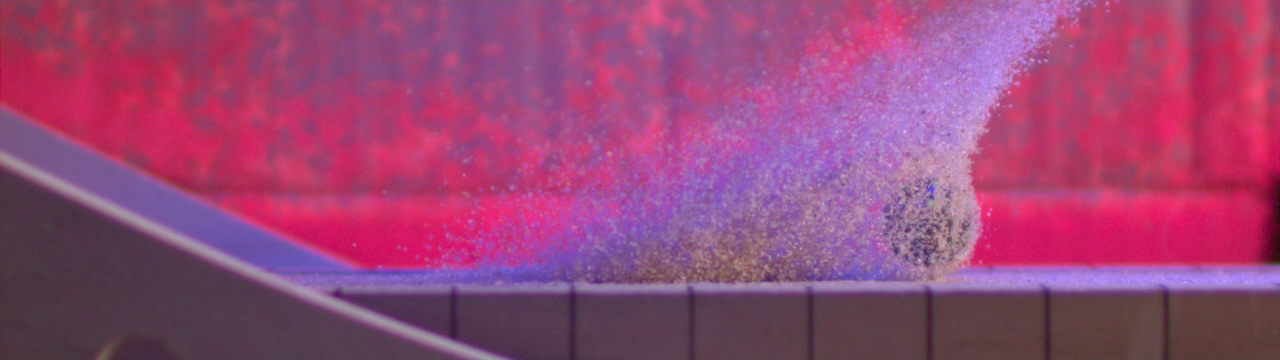}
   \caption{Light sand, \vseventeen}
   \label{fig:impact_sand_light}
\end{subfigure}
\caption{We show the projectile about 66 ms after impact and during its ricochet on \subref{fig:impact_grav_large}) coarse gravel,
\subref{fig:impact_grav_mid}) fine gravel,
\subref{fig:impact_sand_dark}) dark sand, and
\subref{fig:impact_sand_light}) light sand substrates. 
The marble was launched with an impact angle of $30\degree$ and velocities of $\sim 3.8$ m/s in all three experiments. These images are individual frames taken from the three high speed videos detailed in Table \ref{tab:video_list}. 
Ejecta from impacts into gravel have significantly less material compared to sand. The surface of the larger grain substrate is disturbed more compared to the smaller gravel size. The projectile's rebound height is largest for the coarse gravel, our largest grain size.}
\label{fig:ejecta}
\end{figure}

For each of the four substrates we filmed three impact experiments. 
Each impact was filmed simultaneously in high speed using Krontech Chronos 1.4 high speed cameras pointed in two directions.
The first camera filmed a side view in the plane of the pendulum arm (xz-plane). 
Videos filmed from this direction were recorded at a frame rate of 3030 frames per second.
Figure \ref{fig:ejecta} shows snapshots from this camera view about 66 ms after impact for each substrate.
A second high speed camera was used to film the projectile impact in the substrate surface plane (xy-plane).
The pendulum arm makes it impossible to directly film an overhead view of the impact. 
Instead, the camera is positioned at a $45\degree$ angle relative to horizontal and positioned at a height that allows most of the tray length to be visible in each video frame. 
The videos from this elevated view were recorded at a frame rate of 2600 frames per second.
Figure \ref{fig:setup_cartoon} shows a cartoon of the experimental setup and camera positions.
All experiments have approximately the same impact velocity and impact angle. 
The exact values of impact angles and velocities for each experiment are measured from the high speed videos and are listed in Table \ref{tab:video_list}.  
Froude numbers (Equation \ref{eqn:Fr}) computed using the impact velocity for these experiments are also listed in Table \ref{tab:video_list}.
These videos are used to track the position and spin of the marble as detailed in Section \ref{sec:data}.

Before each impact experiment, the substrate is disturbed to reduce local porosity or compaction variations in the substrate, especially near the initial impact site. 
We disturb both sands with a rake of nails spaced one cm apart. 
The rake is run through the length of the substrate tray and the surface is leveled taking care not to compact the sand.
The larger grain size of the gravel made this raking technique ineffective. 
Instead a nail punch was used to disturb the gravel in the tray which then had its surface leveled without pushing down to avoid compaction.

\subsection{Measurement of the deflection angle $\alpha$}
\label{sec:deflection_angle}

With the pre-impact projectile trajectory in the $xz$ plane, we noticed that post-impact trajectories were deflected out of this plane and gained a horizontal velocity component that was particularly noticeable in the coarser media.
We  characterize the deflection with a deflection angle $\alpha$ measured between the tray center-line and the marble's final location in the xy-plane.
In a second set of experiments, 
we filmed about ten impacts into each of the four granular substrates and filmed them with a Nikon V1 camera at a regular frame-rate.
The camera was positioned slightly above and at the end of the tray looking down along its length (x-axis).
See Figure \ref{fig:setup_cartoon} for a cartoon of the Nikon camera position and the deflection angle.

With this camera directed along the length of the tray we are able to measure the $y$ component of the projectile's velocity. Analysis of how the deflection angle is measured from these videos is discussed in Section \ref{sec:deflection_analysis}.

\begin{table}
{
\caption{Projectile and tray properties}
\label{tab:quantities}
\begin{tabular}{lll}
\hline
Radius of marble & $R_p$ & 8.075 mm \\
Mass of marble & $m_p$ &  5.57 g \\
Density of marble & $\rho_p$ & 2.53 g/cm$^3$ \\
Unit of velocity & $\sqrt{g R_p}$ & 28.1 cm/s \\
\hline
Tray 1 dimensions, coarse gravel & 
\multicolumn{2}{c}{ 87.5 x 11.5 x 6.3 cm}\\
Tray 2 dimensions, fine gravel &  
\multicolumn{2}{c}{ 79.9 x 12.9 x 6.2 cm}\\
Tray 3 dimensions, both sands & 
\multicolumn{2}{c}{ 74.9 x 11.1 x 6.3 cm} \\
\hline
\hline
\end{tabular}
}
\end{table}

\begin{table*}
\centering
{
\caption{High Speed video list}
\label{tab:video_list}
\begin{tabular}{llllllllll}
\hline
Video & Material & $\pi_{grain}$ & $v_{impact}$ & $\theta_{impact}$ & Fr \\ 
 & & & (cm/s) & (deg) & \\
\hline
\vthirteen & Gravel, coarse &  & 353 & 29.9 & 12.6\\
\vfourteen & Gravel, coarse & 3.7 & 355 & 29.0 & 12.6\\
\vfifteen & Gravel, coarse &  & 357 & 35.7 & 12.7\\
\hline
\vnineteen & Gravel, fine &  & 359 & 28.9 & 12.8\\
\vtwenty & Gravel, fine & 5.0 & 359 & 30.2 & 12.8\\
\vtwentyone & Gravel, fine &  & 360 & 30.4 & 12.8\\
\hline
\vtwentytwo & Sand, dark &  & 367 & 27.9 & 13.1\\
\vtwentyfour & Sand, dark & 16.5 & 367 & 27.9 & 13.1\\
\vtwentyfive & Sand, dark &  & 362 & 29.7 & 12.9\\
\hline
\vsixteen & Sand, light &  & 359 & 31.0 & 12.8\\
\vseventeen & Sand, light & 50.5 & 387 & 27.9 & 13.8\\
\veightteen & Sand, light &  & 363 & 27.9 & 12.9\\
\hline
\end{tabular}}

Videos listed were recorded from the side view (xz-plane) and recorded at 3030 frames per second.

\end{table*}

\section{Trajectories}
\label{sec:trajectories}

A white light reflection dot on the marble is used to track the trajectory of the marble.
Painted fluorescent green dots on the marble surface are used to track the angular velocity (spin) of the marble.
Data reduction for projectile trajectory and spin tracking is done as described in Section 3.1 by \citet{Wright_2020b} using the python package Trackpy \citep{trackpy}.
We use these trajectories to measure coefficeints of restitutions, deflection angle in the $xy$ plane, a time for the projectile to exit the impact crater $t_e$, and an effective friction coefficient $\mu_{eff}$.

\subsection{Data reduction}
\label{sec:data}

\begin{figure*}
\centering
\begin{subfigure}{\textwidth}
   \centering
   \includegraphics[width=0.9\textwidth,scale=1,trim = {0 0 0 0}, clip]{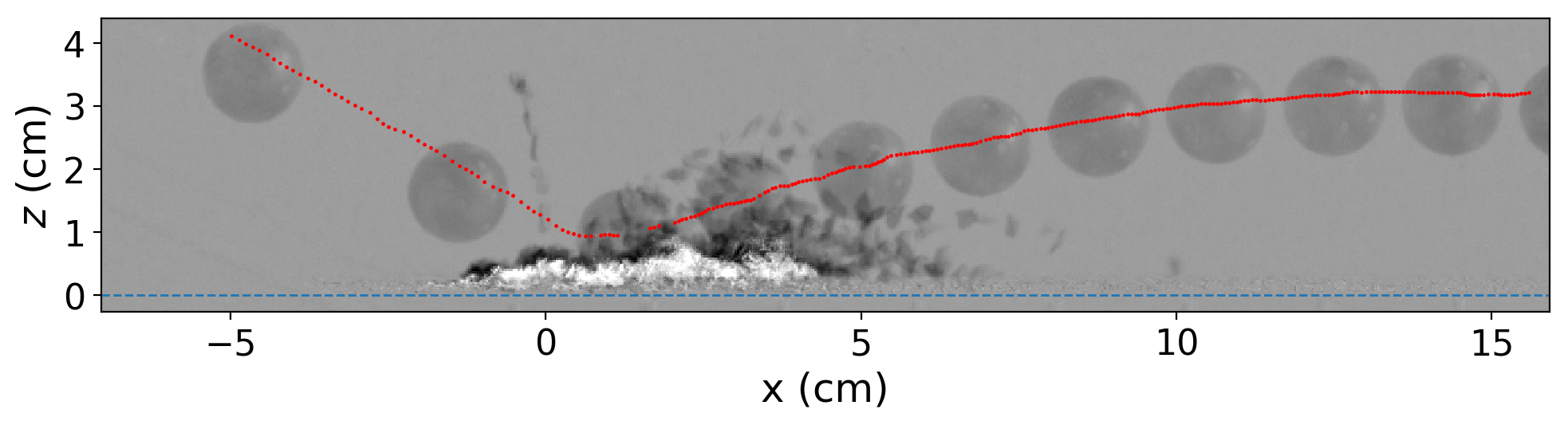}
   \caption{Coarse gravel, \vthirteen}
   \label{fig:grav_coarse_seq} 
\end{subfigure}
\begin{subfigure}{\textwidth}
   \centering
   \includegraphics[width=0.9\textwidth,scale=1,trim = {0 0 0 0}, clip]{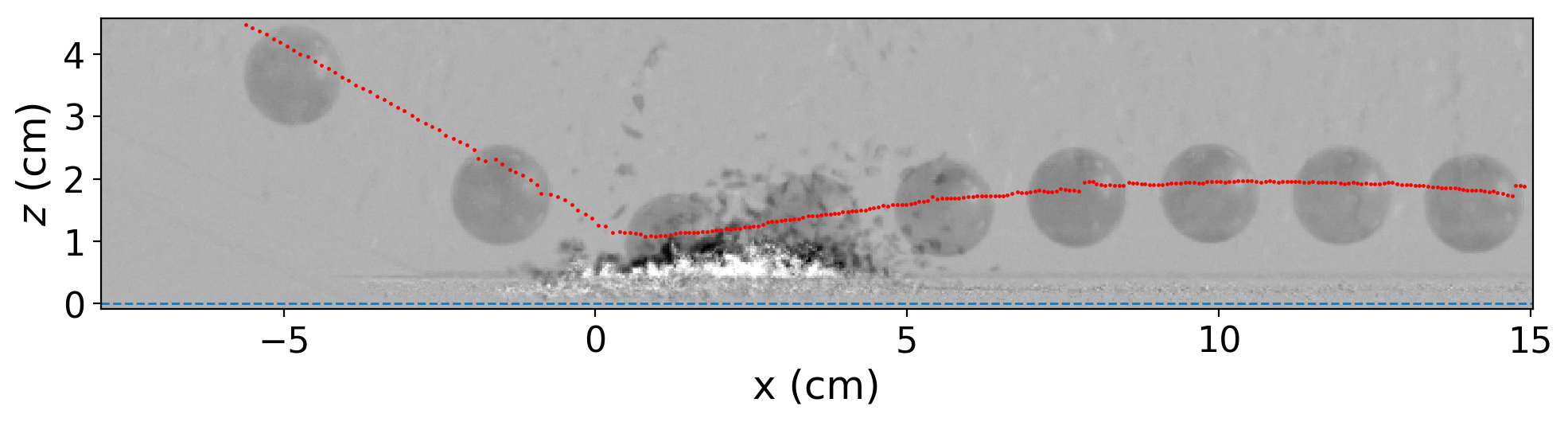}
   \caption{Fine gravel, \vnineteen}
   \label{fig:grav_fine_seq}
\end{subfigure}
\begin{subfigure}{\textwidth}
   \centering
   \includegraphics[width=0.9\textwidth,scale=1,trim = {0 0 0 0}, clip]{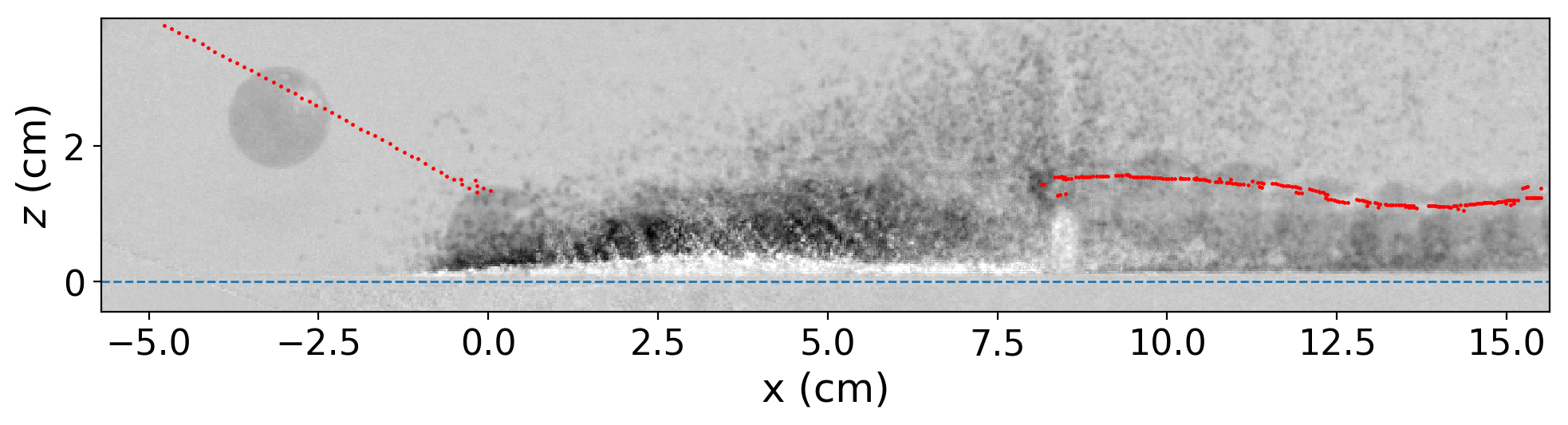}
   \caption{Dark sand, \vtwentytwo}
   \label{fig:sand_dark_seq}
\end{subfigure}
\begin{subfigure}{\textwidth}
   \centering
   \includegraphics[width=0.9\textwidth,scale=1,trim = {0 0 0 0}, clip]{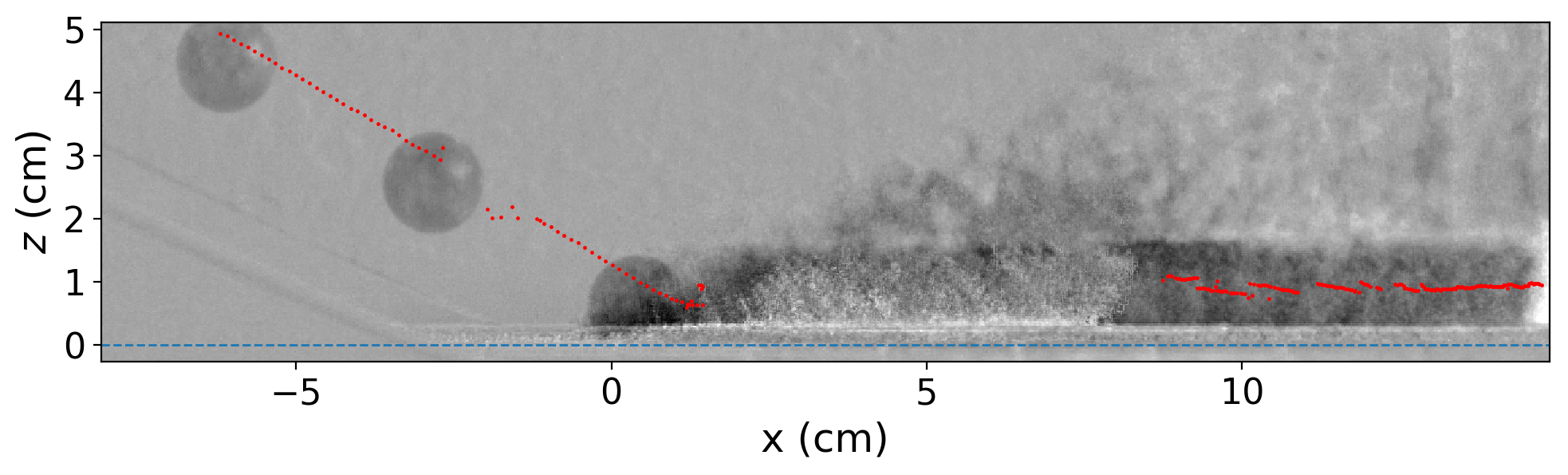}
   \caption{Light sand, \vseventeen}
   \label{fig:sand_light_seq}
\end{subfigure}
\caption{Trajectory of the marble measured by tracking the white light reflection on the marble.
Trajectories are shown for four different high speed videos. From top to bottom, the panels are of experiments launching a marble of radius 8.075 mm into a granular substrate with coarse gravel, fine gravel, dark sand, and light sand.
All four experiments had an impact angle of $30\degree$  and impact velocities of $\sim 3.8$ m/s.
The horizontal and vertical axis are the marble's physical position in centimeters. 
We show in grayscale a sum of high speed video images that are separated by 33 ms time intervals.
The red line is the center of mass track that goes through a white light reflection seen on the left side of the marble. 
The blue dashed line is the granular surface.
The marble started in the upper left corner and came to rest outside the field of view on the right.
With impacts into sand, the marble is partially obscured by the ejecta curtain during the middle part of its trajectory. In all four cases the marble ricocheted, reaching the largest maximum rebound height when impacting the gravel with the largest grain size. 
}
\label{fig:traj_seq}
\end{figure*}

The white light reflection track was measured from the high speed videos of ricocheting impacts on each of the four grain sizes.
Figure \ref{fig:traj_seq} shows an example of the measured tracks of the white dot reflection as red points for each granular size in the pendulum plane.
The tracks are overlayed on a summed image of several frames from the high speed videos spaced by 33 ms.
The dashed blue line denotes the surface of the granular substrate.
The origin is defined as the location of impact, with $x$ increasing to the right along the projectile motion and $z$ increasing in the vertical direction.
All of these experiments had the projectile launched at an angle of $30\degree$ and a velocity of 3.7 to 3.9 m/s.
Information about the high speed videos of our experiments are
given in Table \ref{tab:video_list}.

\begin{figure*}
\centering
\includegraphics[width=\linewidth]{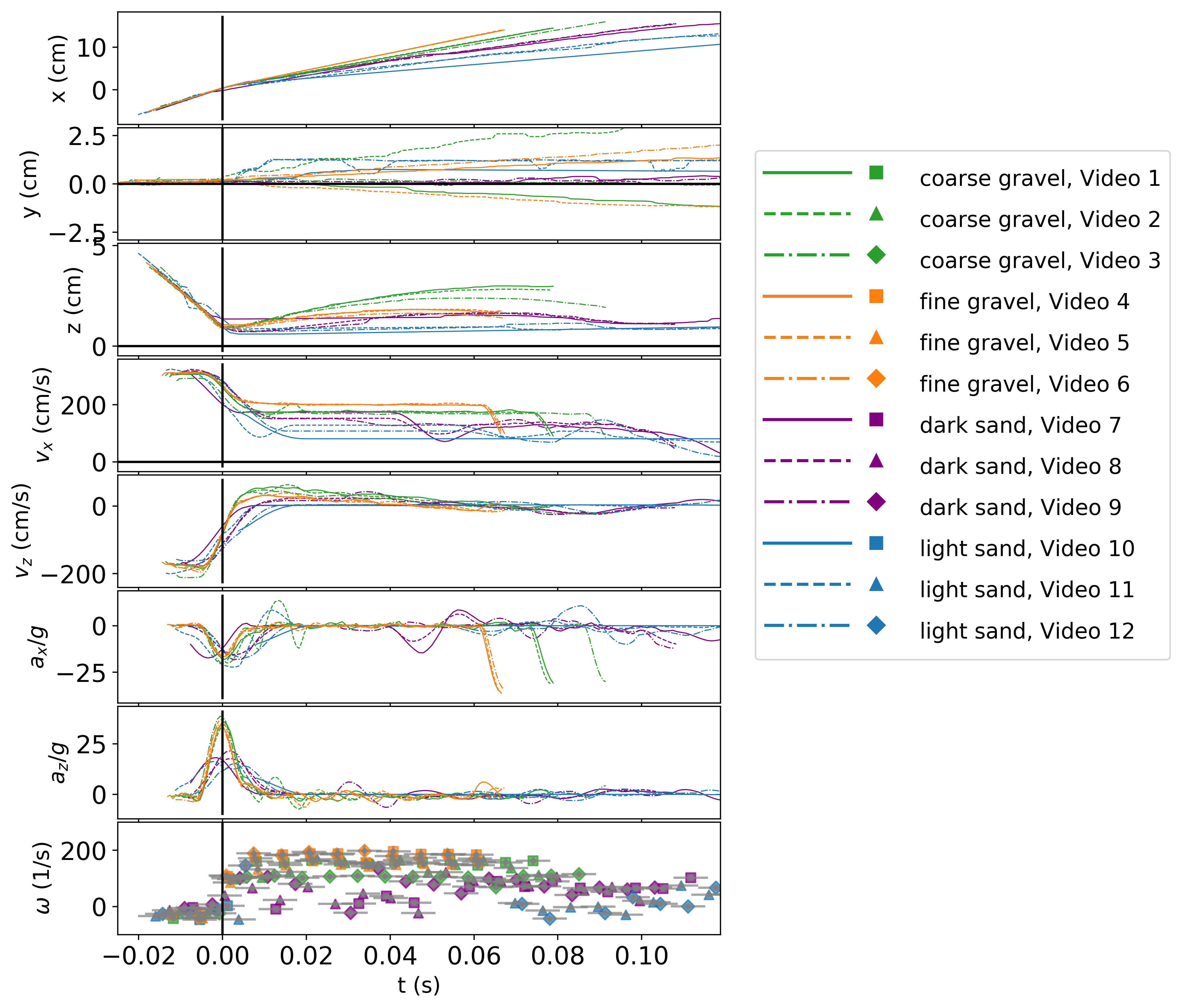}
\caption{Trajectories of the tracked white dot reflection on the marble for 12 impact experiments into various grain sizes. All experiments plotted have an initial impact angle of $30\degree$ and drop height of 110 cm, which corresponds to an impact velocity of 3.7 to 3.9 m/s.
We show the projectile's positions, velocities, accelerations, and spins as a function of time.
A vertical position of $z=0$ is defined at the granular surface. 
The positions $x,y,$ and $z$ are measured from the point of impact with $x$ increasing along the direction of motion (to the right). 
Estimated time of impact ($t=0$) is shown with the black vertical line.
Line color denotes the granular substrate and grain size of the experiment: green is coarse gravel, orange is the finer gravel, purple is the dark sand, and blue is the light sand. 
Line style corresponds to unique experiments for a given granular size and labeled with the video number listed in Table \ref{tab:video_list}.
The bottom panel are the measured angular velocities of the projectile.
Marker color follows the pattern described above, and marker style corresponds to unique experiments of the same grain size.
For the spin in the bottom panel, the horizontal error bars show the time intervals used to measure the spin.
\label{fig:pend_traj}
}
\end{figure*}

The marble's position in each axis, along with the respective velocity and acceleration components, are shown in Figure \ref{fig:pend_traj} as functions of time.
The horizontal $x$ and vertical $z$ positions were measured from a high speed camera looking at the pendulum (xz) plane (see Figure \ref{fig:setup_cartoon}). 
The y-component of the projectile's trajectories were measured from the videos of the surface ($xy$) plane. 
The trajectories were corrected for this camera's $45\degree$ viewing angle relative to horizontal. 
To compute velocities and accelerations from the tracked positions, we smoothed the arrays using a Savinsky-Golay filter. 
In these plots 
$t=0$ corresponds to the time of impact.
For the marble's position plots zero was chosen to be the location of impact for $x$, the surface of the granular substrate for $z$, and the center line of the tray for $y$.

From the $z(t)$ plots it is clear that the rebound height of the trajectories post impact is greatest for the coarse gravel, our largest grain size, and decreases with grain size. 
Consequently, the vertical velocity component is largest for the coarse gravel.
The acceleration is greatest for both gravel sizes suggesting the hydrostatic pressure from the gravel is greatest on coarser material.

The horizontal velocity component $v_x$ was greatest for the fine gravel followed by the coarse gravel and the two sands.
This suggests that the drag force is greatest for the light sand and decreases with the dark sand.
This may be due to the finer gravel forming a smoother surface compared to the coarse gravel allowing the marble to skid across the surface.
Also, the fine gravel has less material ejected after impact compared to all other grain sizes.
More mass in the ejecta curtain being built up in front of the marble could increase the horizontal drag force.

The deflection in the marble trajectory from the tray's center can be seen on the $y(t)$ plot in Figure \ref{fig:pend_traj}.
We found that the coarse gravel has the greatest deflection from center ($y=0$) with the deflection angle $\alpha$ decreasing with smaller grain sizes.
We analyze these deflections in more detail from a second set of experiments described in Section \ref{sec:deflection_angle}.

Fluorescent green dots painted on the surface of the marble were tracked to measure the spin of the marble.
The measured spins are plotted in the bottom cell of Figure \ref{fig:pend_traj}.
Out of all the grain used, the finer gravel produced the largest spin on the projectile.
The coarse gravel and black sand produced comparable spin values with the former being slightly greater.
From inspection of the high speed videos and the measured spins late in time, the marble's spin post impact was less in the light sand than the other three grain sizes.

\begin{table}
{
\caption{Restitution coefficients}
\label{tab:restitution}
\begin{tabular}{l|ll|ll|ll}
\hline
Material & $e_x$ & $\sigma_{e_x}$ & $e_z$ & $\sigma_{e_z}$ & $\mu_{eff}$ & $\sigma_{\mu_{eff}}$\\
\hline
Gravel, coarse\!&\!0.607\!&\!0.015\!&\!0.297\!&\!0.067\!& \!0.115\!&\!0.028\\
Gravel, fine   \!&\!0.680\!&\!0.008\!&\!0.181\!&\!0.021\!& \!0.150\!&\!0.025\\
Sand, dark     \!&\!0.541\!&\!0.113\!&\!0.149\!&\!0.050\!& \!0.067\!&\!0.026\\
Sand, light    \!&\!0.427\!&\!0.068\!&\!0.056\!&\!0.051\!& \!0.107\!&\!0.071\\
\hline
\hline
\end{tabular}
}
\end{table}

\begin{figure}
\centering
\includegraphics[width=\columnwidth]{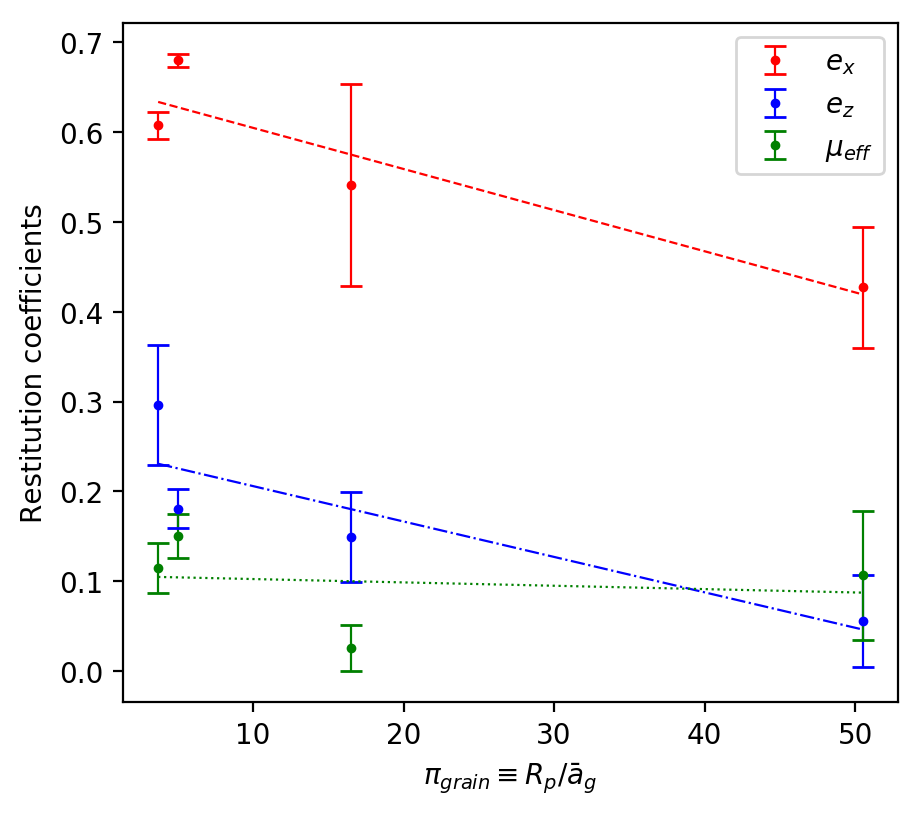}
\caption{
Restitution coefficients with errors of one standard deviation plotted as functions of projectile to mean grain size ratio $\pi_{grain}$.
Larger grain sizes are on the left.  The dashed lines correspond to a linear fit of the coefficients. Values are taken from Table \ref{tab:restitution}. 
Coefficients of restitution are larger in larger grain size substrates.  The effective friction coefficient $\mu_{eff}$ is insensitive to grain size.
\label{fig:restitutions}
}
\end{figure}

\subsection{Restitution coefficients}
\label{sec:restitution}
Using the trajectories from the 12 tracked videos of ricochets discussed in Section \ref{sec:data}, we measured the $x$ and $z$ components of the projectile velocity before and after the impact.
The ratio of the before and after $x$ velocity components we denote $e_x$ and the ratio of the before and after $z$ velocity components we denote $e_z$.  These dimensionless numbers can be described as coefficients of restitution as a perfectly elastic collision would give $e_x = e_z = 1$. 

For each substrate we have three tracked videos (twelve videos in total), which gave us three measurements for the coefficients of restitution, $e_x$ and $e_z$.
The 
three measured values in each substrate for the horizontal coefficient of restitution $e_x$ were used to estimate an uncertainty $\sigma_{e_x}$. We similarly estimated uncertainties $\sigma_z$ for the vertical coefficients of restitution. 
The mean values and uncertainties of the coefficients of restitution $e_x, e_z$ for each substrate are listed in Table \ref{tab:restitution} and are plotted as a function of mean grain size (using $\pi_{grain}$) in Figure \ref{fig:restitutions}.

We computed effective friction coefficients for the spin up caused by frictional contact between the projectile and granular medium $\mu_{eff}$ as described in Section 3.3 by \citep{Wright_2020b}. 
The effective friction coefficient $\mu_{eff} = 2 R_p \dot \omega/ (5a_z)$ (equation 12 by \citealt{Wright_2020b}) is estimated from the measured angular acceleration $\dot \omega$ and vertical acceleration component $a_z$ of the marble projectile.
The mean values and uncertainties of the effective friction coefficients $\mu_{eff}$ for each substrate are also listed in Table \ref{tab:restitution},  and are plotted as a function of mean grain size (using $\pi_{grain}$) in Figure \ref{fig:restitutions}.

Figure \ref{fig:restitutions} shows that the coefficients of restitution $e_x, e_z$ are sensitive to grain size, with larger coefficients at smaller ratio $\pi_{grain}$,  corresponding to larger coefficients with larger grain size.
The $e_z$ coefficient of restitution in the coarse gravel is about five times that in the finer sand.  The $e_x$ coefficient in the coarse gravel is about $35\%$ larger that of the finer sand.  
 
Are these trends significant? 
Previous studies of normal impacts (e.g. \cite{ambroso05}, \cite{goldman08}, \cite{katsuragi07}) have found at most a linear dependence of the empirical force laws on substrate density.  
The differences between substrate mean density in our different media are at most $12.5\%$ (see Table \ref{tab:granular}).
Thus differences in substrate density are unlikely to account for the trends we see in the coefficients of restitution in Figure \ref{fig:restitutions}.

Studies of normal impacts (e.g. \cite{katsuragi13}, \cite{allen57}, \cite{goldman08}) have found at most a linear dependence of the empirical force law coefficients on the 
substrate static friction coefficient. 
Differences in the coefficients of static friction in our substrates are at most $14\%$ (see Table \ref{tab:granular}). 
Thus differences in substrate static friction coefficient are unlikely to account for the trends we see in the coefficients of restitution in Figure \ref{fig:restitutions}.

Figure \ref{fig:restitutions} illustrates that the coefficients of restitution trend toward greater values in substrates with larger mean grain size.  This is consistent with our discussion in Section \ref{sec:data} about the projectile trajectories.  
The dashed lines are linear fits using the least squares method.
The negative linear 
slopes for $e_x$ and $e_z$ are consistent with our observation that the coefficients of restitution are larger in larger grain size substrates. 
Our results are comparable to drop tower experiments by \cite{brisset20} where a projectile normally impacted a granular bed of centimeter sized particles. They found the impacts had a vertical coefficient of restitution of $\sim 0.16$. These are similar to values found for our $e_z$, but unlike \cite{brisset20}, we found that $e_z$ decreases with smaller grain size appreciably as discussed earlier in this section.
Our horizontal coefficient of restitution is much larger than the vertical coefficient for a given grain size suggesting $e_x$ plays a greater role in the dynamics of an oblique impact.

\subsection{The standard deviation of the deflection angle $\alpha$}
\label{sec:deflection_analysis}


From our videos, we measured the deflection angle for each impact using the angle tool in ImageJ which measures the angle of a triangle defined by three points.
We chose our points such that two formed a line from the end of the tray to the marble's impact location. The line formed by these two points is parallel to the tray center-line.
A third point was chosen to be the marble's final location in its trajectory.
The vertex of the triangle was chosen to be the center of the marble at the time and location of impact.
See Figure \ref{fig:setup_cartoon} for a cartoon of the camera positions and deflection angle $\alpha$.
We could have used the 2 cm marks on the sides of the trays to calculate an angle but the distance between markers is distorted due to the camera projection and would introduce errors.

\begin{figure*}
\centering
\includegraphics[width=\columnwidth]{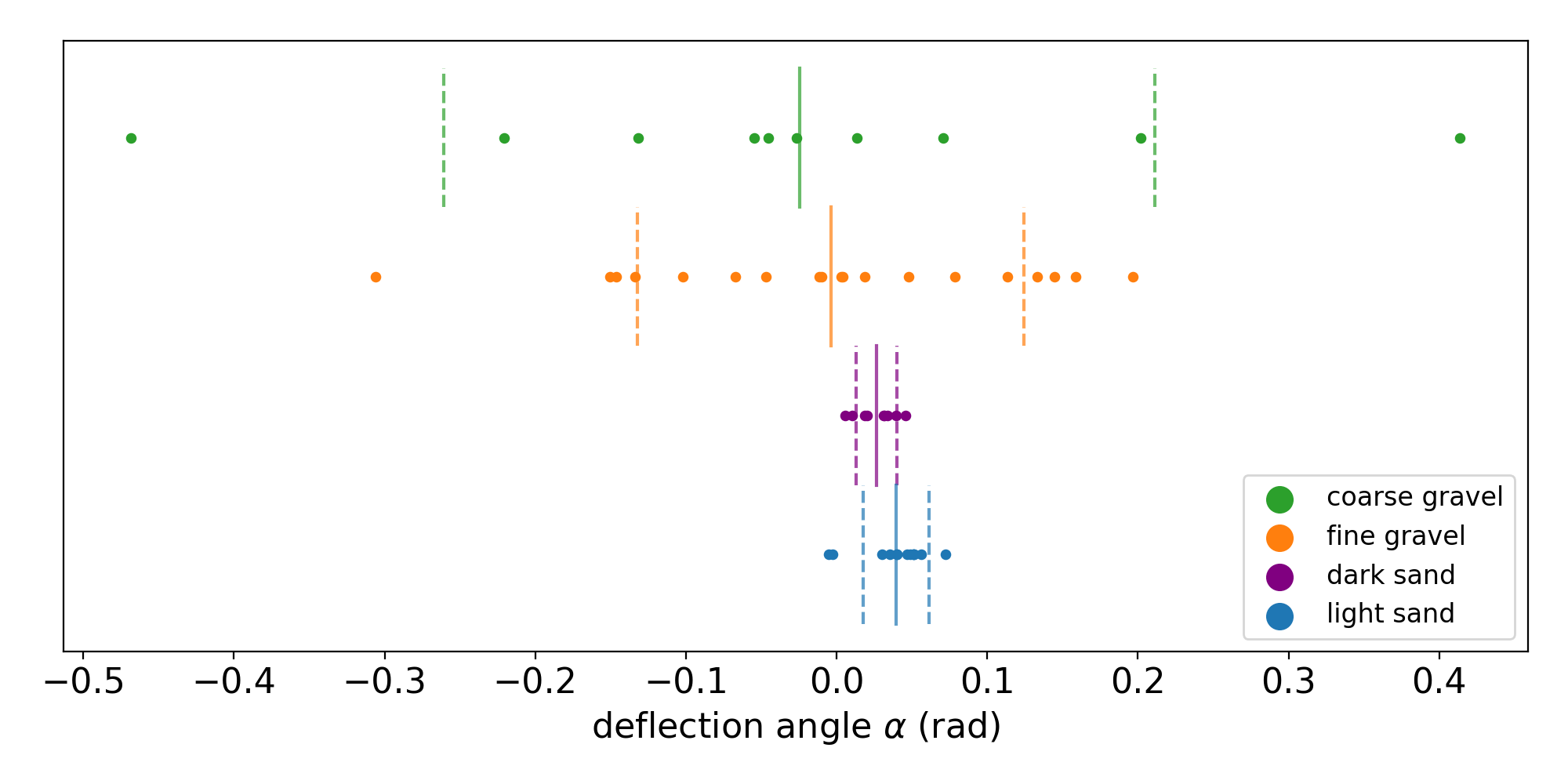}
\caption{
Distribution of deflection angles in the $xy$-plane after impact. Each color is for the different granular substrate sizes with vertical position being an arbitrary offset. The solid and dashed lines are the means and standard deviations, respectively, of their distribution. It is clear that the spread of measured deflection angles is dependent of the size of the grains in the substrate.
\label{fig:ang_dist}
}
\end{figure*}

Multiple impact experiments were done on all four grain sizes with each impact giving a different deflection angle. 
Figure \ref{fig:ang_dist} shows the distribution of measured deflection angles for each granular substrate in its own color. The solid lines are the means of the respective distribution and the dashed lines represent their standard deviations.
We plotted the standard deviations of the measured deflection angles $\sigma_{\alpha}$ for each substrate as shown in Figure \ref{fig:deflection_std}
as a function of projectile to grain size ratio $\pi_{grain}$.  The uncertainty in $\sigma_\alpha$ was estimated by propagating the  errors in our measurements for the deflection angle.

\section{Analysis}

\subsection{A scaling argument for the sensitivity
of the deflection angle to grain size}
\label{sec:deflection_model}

The stopping force of a normal impact was modeled through a collisional-based model 
where momentum transfer from the impactor to the target takes place through sporadic  collisions of high force carrying grains at the intruder surface \citep{bester17}.  We adopt a similar collision-based approach to estimate the dependence of the ricochet deflection angle on grain size.

We would like to find a physical relation between  standard deviation of the deflection angle $\sigma_{\alpha}$ as function of the grain size ratio $\pi_{grain}$.
We assume that the deflection angle $\alpha$ from the center of the tray follows a Gaussian distribution.  We assume  that the interaction of the projectile and granular substrate can be modeled as a random walk process involving multiple collisions with individual grains.
Therefore, the standard deviation of the deflection angle can be written as 
\begin{equation} \label{eq:sig_alpha}
    \sigma_{\alpha} \sim \sqrt{N} S
\end{equation}
where $N$ is the number of steps in the random walk and $S$ is the step size. 
Here $S$ is the change in deflection angle that arises from a typical collision between a single grain and the projectile.

To estimate the number of grain collisions, to order of magnitude, we consider the projectile's momentum to be imparted on $N$ number of grains with momentum $m_g v_g$.
consider the exchange of the projectile's momentum to the total momentum of $N$ number of grains. 
Therefore, we need $N \sim M_p/m_g$ collisions to change the direction of the projectile during the impact.
This is an equivalent argument for calculating the stopping time of the impactor into its target.

The fractional change in projectile momentum direction from one collision scales with $m_g/M_p$ in the center of momentum frame.
Since the collisions can occur anywhere on the projectile's half-sphere we take the change in deflection angle to be of order
$S \approx m_g/M_p$.  This and equation \ref{eq:sig_alpha} gives a standard deviation in the deflection angle  
\begin{equation}
    \sigma_{\alpha} \sim \left( \frac{m_g}{M_p} \right)^{\frac{1}{2}} \sim \left( \frac{\bar a_g}{R_p} \right)^{\frac{3}{2}} \sim \pi_{grain}^{-\frac{3}{2}}. 
\end{equation}

\begin{figure}
\centering
\includegraphics[width=\columnwidth,trim =  0 0 0 0, clip]{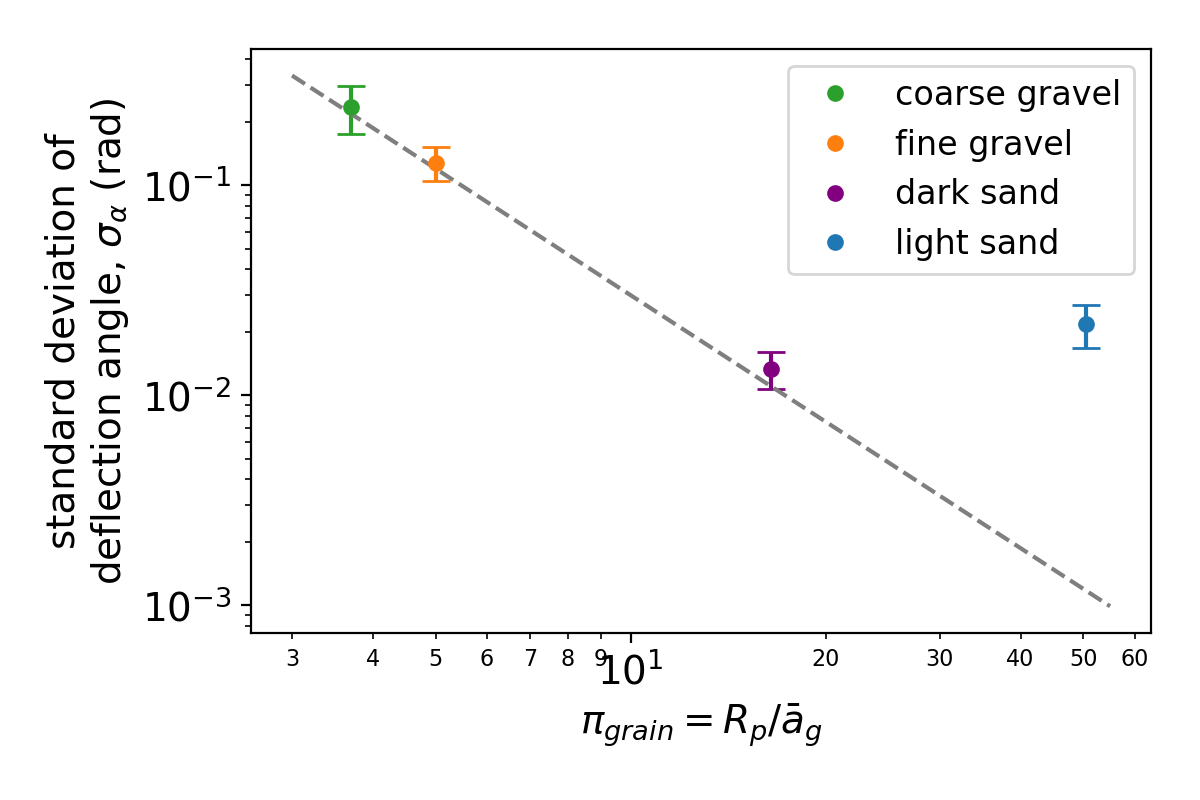}
\caption{Plot of the standard deviation of the deflection angle $\sigma_{\alpha}$ (in radians) of the marble in the horizontal plane versus $\pi_{grain}$, the ratio of projectile to grain length. 
The deflection angle is the deviation of the marble trajectory from the center line of the substrate tray after impact.
Measurements of the deflection angle $\alpha$ in 
several impact experiments were used to calculate the standard deviation of the deflection angle.
The dashed gray line shows the scaling relation 
$\sigma_{\alpha}(\pi_{grain}) = 3 \pi_{grain}^{-3/2}$.
\label{fig:deflection_std}
}
\end{figure}

This simple power law scaling is shown in Figure \ref{fig:deflection_std} as a dashed gray line
of the form $\sigma_{\alpha} = 3 \pi_{grain}^{-3/2}$.
The standard deviation of $\alpha$ for the light colored sand has an unexpectedly large standard 
deviation in the deflection angle 
and is not consistent with our simple power law scaling.
Instead there appears to be asymptotic behavior in  $\sigma_\alpha$ with decreasing grain size.

It was shown in \cite{quillen19} that the permeability of a granular media can result in an aerodynamic force that causes a suction on a very flat object that is rising off the surface.
The flow of air in the granular media could change the forces acting on the marble during the penetration and rebound phases of its trajectory.
The permeability of a granular system is proportional to the average flow velocity in the material.
This velocity is dependent on the porosity of the material with smaller values constricting the fluid movement.
Consequently, the permeability $\kappa \sim \phi^{-1}$ would be highest for the light playground sand and air flow in the medium could be important to the forces during and after the impact for this grain size.

\subsection{Phenomenological model of lift and drag coefficients $C_L$ and $C_{Dx}$}

Phenomenological models have been developed for laboratory studies of spherical projectiles impacting into granular substrates at normal incidence  (e.g. \cite{ambroso05}, \cite{tsimring05}, \cite{katsuragi07}, \cite{katsuragi13}, \cite{altshuler14}, \cite{murdoch17}).
The projectile's equation of motion is usually described as
\begin{equation}
    \frac{d^2z}{dt^2} = \frac{dv_z}{dt} = -g + \frac{F_d}{M_p}
\label{eq:eq_motion}
\end{equation}
where $z$ is the vertical coordinate with $z=0$ being the point of impact and positive above the granular surface. $M_p$ is the mass of the projectile, $g$ is magnitude of the acceleration due to gravity, $v_z$ is the projectile's vertical velocity, and $F_d$ is a vertical force from the granular substrate that acts to decelerate the projectile.
A form for $F_d$ often adopted  
\begin{equation}
    F_d = F_z(z) + \alpha_D v_z^2
\label{eq:vertical_force}
\end{equation}
includes a hydrostatic, depth dependent term $F_z(z)$ \citep{katsuragi07} and a hydrodynamic-like drag force (e.g. \cite{allen57}, \cite{tsimring05}, \cite{katsuragi07}, \cite{goldman08}, \cite{pachecovazquez11}, \cite{murdoch17}).
The drag term is attributed to momentum transfer of projectile-grain and grain-grain collisions during impact.

We explore modifications to these phenomenological models that can account for the sensitivity of our measured coefficients of restitution to grain size.  
\cite{Wright_2020b} used a force law similar to Equation \ref{eq:eq_motion} in both the horizontal and vertical directions  to delineate the behavior of a projectile's oblique impact into sand, either ricocheting, rolling out, or stopping in its initial crater.  We adopt a similar model but do not adopt separate force laws for  penetration and rebound phases of the impact. We note that this model is the simplest one that gives both horizontal and vertical coefficients of restitution.

The projectile's horizontal force is modeled with a drag like force
\begin{equation}
    \frac{dv_x}{dt} = - \alpha_x v_x^2
\label{eq:eom_x}
\end{equation}
with coefficient $\alpha_x$ having units of inverse length and $v_x >0$.
For the vertical acceleration we use a model that has a lift force, similar in form to a drag force, that is dependent on the horizontal velocity component 
\begin{equation}
    \frac{dv_z}{dt} = \alpha_L v_x^2 + \beta_z.
\label{eq:eom_z}
\end{equation}
The term $\beta_z$ has units of acceleration and includes hydrostatic and gravitational accelerations which we assume to be constant. 

It is convenient to write our drag and lift parameters in terms of dimensionless coefficients. The drag like parameter $\alpha_x$ can be written
\begin{equation} \label{eq:ax}
    \alpha_x = C_{Dx} \frac{\rho_s A_p}{M_p}
\end{equation}
where $C_{Dx}$ is a dimensionless drag coefficient, $M_p$ is the projectile's mass, and $A_p$ is the cross-sectional area.
A similar relation can be made with our lift parameter
\begin{equation} \label{eq:aL}
    \alpha_L = C_L \frac{\rho_s A_p}{M_p}
\end{equation}
where $C_L$ is a dimensionless lift coefficient.
The constant acceleration term $\beta_z$ can be written in terms of a dimensionless constant $C_z$
\begin{equation}
    \beta_z \equiv C_z \frac{\rho_s R_p^3}{M_p} g
\end{equation}

Equation \ref{eq:eom_x} has the solution

\begin{equation}
    v_x(t) = \frac{v_{x0}}{\alpha_x v_{x0} t + 1}
\label{eq:vx_solu}
\end{equation}
where the initial horizontal velocity component is $v_{x0} = v_x(0)$.
We can relate our impact velocity to the initial velocity components and impact angle
\begin{align}
    v_{x0} = v_{imp} \cos{\theta_{imp}} \\
    v_{z0} = v_{imp} \sin{\theta_{imp}} .
\end{align}

We integrate Equation \ref{eq:eom_z} and substitute Equation \ref{eq:vx_solu} to get the horizontal velocity $v_z(t)$,
\begin{equation}
    v_z(t) \approx \frac{\alpha_L v_{x0}^2 t}{(\alpha_x v_{x0} t + 1)^2} - |v_{z0}| + \beta_z t.
\end{equation}
We integrate $v_z(t)$
\begin{equation}
    z(t) = \frac{\alpha_L}{\alpha_x} v_{x0} \left[t - \frac{\log(\alpha_x v_{x0} t + 1 ))}{\alpha_x v_{x0}} \right] - |v_{z0}|t + \frac{1}{2} \beta_z t^2.
\label{eq:zt}
\end{equation}
to get the particle's horizontal position in time.
We can define an exit time $t_e$ as the time it takes for the projectile to exit the medium at $z(t_e) = 0$. We solve Equation \ref{eq:zt} for this exit time giving
\begin{equation} \label{eq:te}
    t_e \approx \frac{2 |v_{z0}|}{\alpha_L v_{x0}^2 + \beta_z}
\end{equation}
By measuring the exit time of the projectile for different substrate grain size we can infer the value of $\alpha_L$ and its dependence on substrate grain size.

The horizontal coefficient of restitution is given by
\begin{equation} \label{eq:rest_x}
\begin{split}
    e_x \equiv \frac{v_x(t_e)}{v_{x0}} &= \frac{1}{v_{x0} \alpha_x t_e + 1} \\
    &= \frac{1 + \frac{\beta_z}{\alpha_x v_{x0}^2}}{1 + \frac{\beta_z}{\alpha_x v_{x0}^2} + 2\frac{\alpha_x}{\alpha_L}\frac{|v_{z0}|}{v_{x0}}} \\
    &= \frac{1 + \pi_\beta}{1 + \pi_\beta + \pi_\alpha}
\end{split},
\end{equation}
where $\pi_\alpha$ and $\pi_\beta$ are dimensionless constants given by
\begin{equation}
    \pi_{\alpha} \equiv 2 \frac{\alpha_x}{\alpha_L}\frac{|v_{z0}|}{v_{x0}}
    =2 \frac{C_{Dx}}{C_L} \tan \theta_{imp}
\end{equation}
and
\begin{equation}
\begin{split}
    \pi_{\beta} \equiv \frac{\beta_z}{\alpha_x v_{x0}^2} = \frac{C_z}{C_{Dx}}\frac{g R_p}{v_{imp}^2}\frac{1}{\cos^2\theta_{imp}}.
\end{split}
\end{equation}

Similarly, the vertical coefficient of restitution is
\begin{equation} \label{eq:rest_z}
\begin{split}
    e_z \equiv \frac{v_z(t_e)}{v_{z0}}
    &= \frac{2}{1 + \frac{\beta_z}{\alpha_x v_{x0}^2} + 2\frac{\alpha_x}{\alpha_L}\frac{|v_{z0}|}{v_{x0}}} + \frac{\frac{\beta_z}{\alpha_x v_{x0}^2}-1}{\frac{\beta_z}{\alpha_x v_{x0}^2}+1}\\
    &= \frac{2}{1 + \pi_\beta + \pi_\alpha} + \frac{\pi_\beta - 1}{\pi_\beta+1}.
\end{split}
\end{equation}

Using our definition of the Froude number $Fr=v_{imp}/\sqrt{gR_p}$, we can write $\pi_\beta$ as
\begin{equation}
\begin{split}
    \pi_{\beta} \equiv \frac{C_z}{C_{Dx}}\frac{1}{Fr^2 \cos^2\theta_{imp}}
\end{split}
\end{equation}
The Froude numbers for our experiments were $Fr \sim 13$ for all grain sizes and this implies that $\pi_\beta \ll 1$.

In the high Froude number limit, the restitution coefficients simplify to
\begin{align}
    e_x \sim \frac{1}{1+\pi_\alpha} \\
    e_z \sim \frac{1-\pi_\alpha}{1+\pi_\alpha}.
\end{align}
At high velocities the coefficients of restitution are unaffected by the constant acceleration term $\beta_z$ that was included in Equation \ref{eq:eom_z}.
By taking the ratio of restitution coefficients we can estimate the ratio of drag $C_{Dx}$ to lift $C_L$ coefficients
\begin{align} 
    \pi_\alpha = 1- \frac{e_z}{e_x} = 2\frac{C_{Dx}}{C_L}\tan\theta_{imp}\\
    \frac{C_{Dx}}{C_L} = \frac{1}{2\tan\theta_{imp}} \left( 1- \frac{e_z}{e_x} \right).
\label{eq:ratioC}
\end{align}

From our high speed videos we measure exit time $t_e$ for each grain size. We find that the largest grain size (coarse gravel) medium has the shortest exit time, with the smallest grain size (light sand) medium having the longest time.
Measured values for $t_e$ in each substrate are given in Table \ref{tab:model_values}.

\begin{figure}
\centering
\includegraphics[width=\columnwidth]{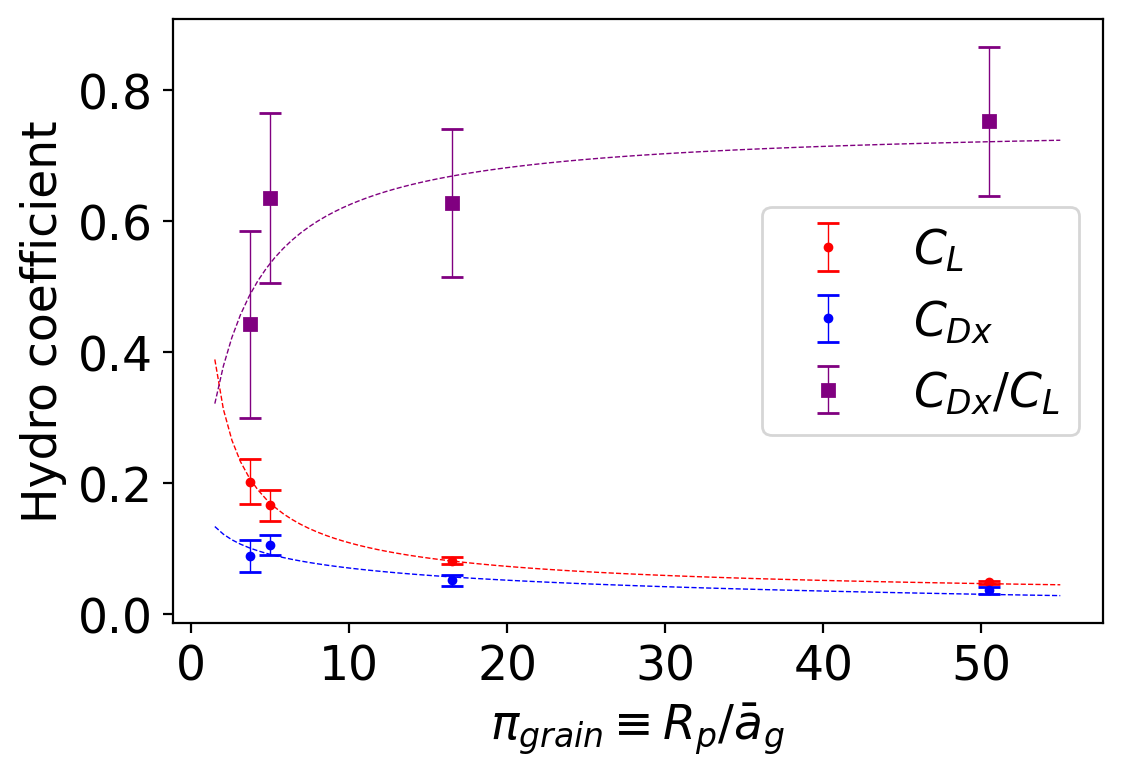}
\caption{Lift and drag coefficients as a function of size ratio $\pi_{grain}$ from oblique impact model. Red dots are the lift coefficients given by Equation \ref{eq:aL}. The drag coefficients are in blue dots calculated using Equation \ref{eq:ratioC}. The ratio of the two coefficients are shown in purple squares.
Dotted lines correspond to fitted lines for the drag, lift, and ratio of the coefficients 
given by Equations \ref{eq:cD_fit}, \ref{eq:cL_fit}, and \ref{eq:ratio_fit}, respectively.
\label{fig:hydro_coeff}
}
\end{figure}

We use our measured values for exit time $t_e$ and Equations \ref{eq:te}, \ref{eq:aL}, and \ref{eq:ratioC} to compute the horizontal drag coefficient $C_{Dx}$ and the vertical lift coefficient $C_L$.
Measured values of the coefficients for each grain size are listed in Table \ref{tab:model_values}. The coefficients are plotted in Figure \ref{fig:hydro_coeff} as a function of size ratio $\pi_{grain}$.
In the figure, the blue dots correspond to the horizontal drag coefficient $C_{Dx}$ and the red dots are for the vertical lift coefficient $C_L$.
The purple squares denote the values of the ratio of the coefficients $C_{Dx}/C_L$.

We found that the largest grain size medium (small $\pi_{grain}$) has the largest lift coefficient with progressively smaller grain size media having smaller lift  coefficients.
This result from the model agrees with measured horizontal positions shown in Figure \ref{fig:pend_traj}.
The drag coefficients are largest for the largest grain sizes and decrease with smaller grains.
This is in agreement from inspection of the impact ejecta curtains. 
Smaller grain sizes had much more material launched by the impact implying more momentum was transferred to launching smaller sized material compared to the larger grain sizes.

Dotted lines in Figure \ref{fig:hydro_coeff} are power law fits we applied to our values for the lift and drag coefficients
\begin{align} \label{eq:cD_fit}
\begin{split}
    C_{Dx} &\approx \frac{1}{4} \pi_{grain}^{-1/6} - 0.08\\
    &= \frac{1}{4} \left( \frac{\bar a_g}{R_p} \right)^{1/6} - 0.08\\
\end{split}
\end{align}
\begin{align} \label{eq:cL_fit}
\begin{split}
    C_L &\approx \frac{1}{2} \pi_{grain}^{-3/4} + 0.02\\
    &= \frac{1}{2} \left( \frac{\bar a_g}{R_p} \right)^{3/4} + 0.02.
\end{split}
\end{align}
Here the drag and lift coefficients have a dependence on grain size to the $1/6$ and $3/4$ power, respectively.
We found that the ratio of the drag to lift coefficients followed
\begin{equation} \label{eq:ratio_fit}
    \frac{C_{Dx}}{C_L} \approx  \frac{3}{4} \left( \frac{\pi_{grain}}{2 + \pi_{grain}} \right)
\end{equation}
which has a form similar to some inertial regime changes described in the literature (e.g., \citealt{Jop_2006}).

We find that the contact time for the projectile in the granular medium is primarily dependent on the strength of the lift force.
A larger lift coefficient decreases the time the projectile is in contact with the granular substrate, increasing the vertical and horizontal coefficients of restitution.
From the simple empirical force model we find that the coefficients primarily depended on the ratio of the inertial drag and lift coefficients.
Using our measured coefficients of restitution, we estimate that ratio of the inertial drag and lift coefficient varies from about 0.4 to about 0.6 for projectile to grain size ratio $\pi_{grain}$ ranging from 4 to 51.

\begin{table}
{
\caption{Values from Phenomenological Model}
\label{tab:model_values}
\begin{tabular}{lllll}
\hline
Material & $t_e$ (ms) & $C_L$ & $C_{Dx}$ & $C_{Dx}/C_L$\\
\hline
Gravel, coarse & 10 & 0.20 & 0.09 & 0.44 \\
Gravel, fine & 12 & 0.17 & 0.11 & 0.64 \\
Sand, dark & 25 & 0.08 & 0.05 & 0.63 \\
Sand, light & 40 & 0.05 & 0.04 & 0.75 \\
\hline
\end{tabular}} 

Note: $\sigma_{t_e} \sim$ 1 ms for all materials and videos 
\end{table}

\section{Summary and Discussion}

We have presented a comparative study of low velocity grazing impacts into four different granular media.  We carried out a series of experiments with the same projectile, impact velocity and angle. The granular media are comprised of similar materials and have similar bulk density, porosity, and friction coefficient, however the mean grain sizes differ (characterized by the dimensionless number $\pi_{grain}$). The spherical projectile rebounds or ricochets off the granular substrate in all the experiments.  
We measure  coefficients of restitution or the ratio of vertical velocity components before and after the impact and the ratio of horizontal velocity components before and after the impact.
We find that the coefficients of restitution are sensitive to grain size with finer granular media having lower coefficients of restitution.  We interpret this result to imply that coefficients of empirical force laws (for hydro-static-like, drag-like and lift-like forces) are sensitive to mean grain size.  

To test our interpretation of the dimensional empirical force law coefficients we developed a phenomenological model to relate them to dimensionless lift and drag hydrodynamic coefficients $C_L$ and $C_{Dx}$, respectively.
We relate the dimensional coefficients $\alpha_L$ and $\alpha_x$ to the dimensionless coefficients with the time it takes the projectile to impact and then exit the granular media.
This exit time $t_e$ was measured for each substrate with the smallest grain size having an exit time of about four times longer than the coarsest grain size.
The coefficient that is most strongly sensitive to grain size is the lift coefficient $C_L$ that decreases by a factor of four between our coarsest and finest media.
The drag coefficient by comparison only varied by a factor of two.

In our experiments we also measure the deflection angle of the projectile, which is is the angle between pre- and post ricochet velocity vectors. 
The deflection angles are largest in the coarser media and their standard deviation approximately depends on the grain's semi-major axis to the 3/2 power.  
This scaling is matched with a model where momentum transfer takes place via collisions with individual grains. 

At lower $\pi_{grain}$, dynamics is essentially only impacts between two objects, the projectile and a single grain.  At higher $\pi_{grain}$ we expect dynamics to become independent of grain size as the grains are small compared to the projectile and act more fluid like.
The range of projectile to grain sizes ratio of our experiments covers a range where the impacts dynamics would be sensitive to grain size.
Future work can explore a larger range of size ratios and more polydisperse substrates with broader axes distributions.
Furthermore, changing the impactor geometry may alter the dynamics of an oblique impact for a given impactor-grain size ratio.
The dependence of impact mechanics on the mean substrate particle size should be considered in future impact models for populations of objects that impact granular asteroid surfaces.

\vskip 1 truein 
\textbf{Acknowledgements}

This material is based upon work supported in part by NASA grants 80NSSC21K0143 and 80NSSC17K0771. 

We thank Jim Alkins for helpful discussions regarding machining.
We are grateful to Tony Dimino for lent equipment.
This study was initiated in collaboration with Randal C. Nelson\footnote{\url{https://www.rochester.edu/newscenter/remembering-randal-nelson-computer-science-professor-428482/}
} 
who is sorely missed.

\vskip 0.3 truein


\bibliographystyle{elsarticle-harv}
\bibliography{ric_v4}





\end{document}